\documentclass[preprintnumbers, floatfix, letterpaper, twocolumn,aps,prd,epsfig,nofootinbib,natbib,longbibliography]{revtex4-1}
\usepackage{graphicx}
\usepackage{epstopdf}
\usepackage{latexsym}
\usepackage{amssymb}
\usepackage{amsmath}
\usepackage{color}
\usepackage{mathrsfs}
\usepackage{xparse}
\usepackage{bbding}
\usepackage{pifont}
\usepackage{comment}
\usepackage{ulem}
\usepackage{float}
\usepackage[inline]{enumitem}
\usepackage{soul}
\usepackage[caption=false]{subfig}

\let\Right\right
\let\Left\left
\makeatletter
\def\right#1{\Right#1\@ifnextchar){\!\right}{}}
\def\left#1{\Left#1\@ifnextchar({\!\left}{}}
\makeatother

\usepackage[
            pdfstartview=FitH,
            bookmarksnumbered=true,
            bookmarksopen=true,
            colorlinks,
            linkcolor=blue,
            anchorcolor=green,
            citecolor=blue
            ]{hyperref}
\begin{document}

  \renewcommand\arraystretch{2}
 \newcommand{\bq}{\begin{equation}}
 \newcommand{\eq}{\end{equation}}
 \newcommand{\bqn}{\begin{eqnarray}}
 \newcommand{\eqn}{\end{eqnarray}}
 \newcommand{\nb}{\nonumber}
 \newcommand{\lb}{\label}
 
\newcommand{\La}{\Lambda}
\newcommand{\va}{\scriptscriptstyle}
\newcommand{\be}{\nopagebreak[3]\begin{equation}}
\newcommand{\ee}{\end{equation}}

\newcommand{\ba}{\nopagebreak[3]\begin{eqnarray}}
\newcommand{\ea}{\end{eqnarray}}

\newcommand{\la}{\label}
\newcommand{\n}{\nonumber}
\newcommand{\su}{\mathfrak{su}}
\newcommand{\SU}{\mathrm{SU}}
\newcommand{\U}{\mathrm{U}}
\newcommand{\red}{ }

\newcommand{\R}{\mathbb{R}}

 \newcommand{\cb}{\color{blue}}
    \newcommand{\cc}{\color{cyan}}
        \newcommand{\cm}{\color{magenta}}
\newcommand{\rc}{\rho^{\scriptscriptstyle{\mathrm{I}}}_c}
\newcommand{\rd}{\rho^{\scriptscriptstyle{\mathrm{II}}}_c} 
\NewDocumentCommand{\evalat}{sO{\big}mm}{%
  \IfBooleanTF{#1}
   {\mleft. #3 \mright|_{#4}}
   {#3#2|_{#4}}%
}
\newcommand{\PRL}{Phys. Rev. Lett.}
\newcommand{\PL}{Phys. Lett.}
\newcommand{\PR}{Phys. Rev.}
\newcommand{\CQG}{Class. Quantum Grav.}

\title{Universal horizons and black hole spectroscopy in  gravitational theories with broken Lorentz symmetry}

\author{Chao Zhang${}^{a, b, c, d}$}
\email{  {chao123@zjut.edu.cn; a30165@rs.tus.ac.jp}}
\author{Anzhong Wang${}^{e}$\footnote{Corresponding author}}
\email{Anzhong$\_$Wang@baylor.edu; Corresponding author}
 \author{Tao Zhu${}^{a, b}$}
\email{zhut05@zjut.edu.cn}
\affiliation{
${}^{a}$ Institute for theoretical physics and cosmology, Zhejiang University of Technology, Hangzhou, 310023, China\\
${}^{b}$ United Center for Gravitational Wave Physics (UCGWP), Zhejiang University of Technology, Hangzhou, 310023, China\\
${}^{c}$ College of Information Engineering, Zhejiang University of Technology, Hangzhou, 310023, China \\
${}^{d}$ Department of Physics, Faculty of Science, Tokyo University of Science,
1-3, Kagurazaka, Shinjuku-ku, Tokyo 162-8601, Japan\\
${}^{e}$GCAP-CASPER, Physics Department, Baylor University, Waco, Texas, 76798-7316, USA}

\date{\today}

\begin{abstract}

The violation of Lorentz invariance (LI) in gravitational theories, which allows superluminal propagations, dramatically alters the causal structure of the spacetime and modifies the notion of black holes (BHs). Instead of metric horizons, now  universal horizons (UHs) define the boundaries of BHs, within which a particle cannot escape to spatial infinities even with an infinitely large speed. Then, a natural question is  how the quasi-normal modes (QNMs) of a BH are modified, if one considers the UH as its causal boundary. In this paper, we study in detail this problem in Einstein-aether theory, a vector-tensor theory that  violates LI but yet is self-consistent and satisfies all observations to date. Technically, this poses several challenges, including singularities of the perturbation equations across metric horizons and proper identifications of ingoing modes at UHs. After overcoming these difficulties, we show that the QNMs  of the Schwarzschild BH, also a solution of Einstein-aether theory, consist of two parts, the metric and aether parts. The QNMs of the metric perturbations are quite similar to those obtained in general relativity and are consistent with current observations of gravitational waves. But the ones from aether perturbations are different, and our numerical studies indicate that they are even not stable. The latter is consistent with our previous studies, which showed that the stealth Schwarzschild BH suffers a Laplacian instability along the angular direction. The method and techniques developed in this paper can be applied to the studies of QNMs in other theories of gravity with broken LI.

\end{abstract}


\maketitle
\section{Introduction}

\renewcommand{\theequation}{1.\arabic{equation}} \setcounter{equation}{0}

The detection of the first gravitational wave (GW) from the coalescence of two massive black holes (BHs) by advanced LIGO marked the beginning of a new era --- {\it the GW astronomy}  \cite{Ref1}. Following this observation, about 90 GW events have been identified by the LIGO/Virgo/KAGRA (LVK) scientific collaborations (see, e.g., \cite{GWs,GWs19a,GWs19b,GWsO3b}). In the future, more ground- and space-based GW detectors will be constructed  \cite{Moore2015,Gong:2021gvw}, which will enable us to probe signals with a much wider frequency band and   {larger distances}. This triggered the interests on the quasi-normal mode (QNM) of black holes, as GWs emitted in the ringdown phase can be considered as the linear combination of these individual modes \cite{Berti2009,Berti18}.

From the classical point of view, QNMs of black holes are eigenmodes of dissipative systems. The information contained in QNMs provide the keys in revealing whether  BHs are ubiquitous in our universe, and more important whether general relativity (GR) is the correct theory to describe gravity even in the strong field regime  \cite{Berti18b}. Basically, a QNM frequency $\omega$ contains two parts, the real and imaginary parts. Its real part gives the frequency of vibration while its imaginary part provides the damping time.  

In  GR,  according to the no-hair theorem, an isolated and stationary BH is completely characterized by only three quantities, mass,  angular momentum and electric charge. Astrophysically, we expect BHs to be neutral,   {so they are} uniquely  described by the Kerr solution. Then, the QNM frequencies and damping times will depend only on the mass and angular momentum of the finally formed BH. Clearly, to extract physics from the ringdown phase, at least two QNMs
are needed. This will require the signal-to-noise ratio (SNR) to be of the order 100  \cite{Berti18}. Although such high SNRs are not achievable right now, it has been shown that they may be achievable once the advanced LIGO, Virgo and KAGRA reach  their fully designed sensitivities. In any case, it is certain that they will be detected by the ground-based third-generation detectors, such as Cosmic Explorer \cite{CE} and the Einstein Telescope \cite{ET}, as well as the space-based detectors, including LISA \cite{LISA}, TianQin \cite{Liu2020}, Taiji \cite{Taiji2}, and DECIGO \cite{DECIGO}.

QNMs in GR have been studied extensively \cite{Chandra92}, including  scalar, vector and tensor perturbations \cite{Iyer1987}. Such calculations have been extended from the Schwarzschild BH to other more general cases, e.g., the Kerr BH \cite{Det1980, Seidel1990}. In this procedure, several different techniques of computations of QNMs were developed. For instance, the  Wentzel-Kramers-Brillouin (WKB) approach \cite{Will1985, Will1987, Konoplya2003, Jerzy2017}, the finite difference method (FDM) \cite{XinLi2020}, the continued fraction method \cite{Leaver1985}, the shooting method \cite{Chandra1975, Doneva2010}, the matrix method \cite{Kai2017}, and so on \cite{Kono2011, Gund1994, Bin2004}. Some of these methods have been also applied to
 modified theories of gravity \cite{Xin2020, Oliver2019}. Additionally, some special approximations, e.g., the eikonal limit,    {have also been} extensively explored,
  see, for example, \cite{Huan2012} and references therein.

In this paper, we shall focus on the QNMs  of black holes in Einstein-aether theory ($\ae$-theory) \cite{JM01}. Such studies are well motivated. In particular, the theory   is self-consistent, such as free of ghosts and instability \cite{Jacobson},
and satisfies all the experimental tests carried out so far \cite{OMW18}. Its Cauchy problem is also well posed \cite{SBP19}, and energy is always positive
(as far as the hypersurface-orthogonal aether field is concerned) \cite{GJ11}.
In addition, there exist both gravitational waves \cite{Foster06,Foster07,Yagi:2013qpa,Yagi14,HYY15,GHLP18,Kai19,Zhao19,Chao2020,OMW21,GHBBCYY} and BHs \cite{Eling2006-1,Eling2006-2,Tamaki2008,BS11,BJS11,Per12,Wang13,Dingq15,Chao2020b,AFJW21,CSS21}.
It was also shown that universal horizons  can be formed from gravitational collapse of realistic matter \cite{BMSW18}, in addition to the formation of the spin-0 and metric
horizons \cite{Garfinkle2007}.

In comparison with other modified theories of gravity \cite{CFPS12}, including  scalar-tensor theories and their high-order corrections \cite{DL19}, $\ae$-theory has at least two distinguishable features:

\begin{itemize}

\item  It is a particular  vector-tensor theory in which the vector field is always timelike. As a result, it always defines a preferred frame and whereby violates locally the Lorentz invariance (LI).  Despite  the facts that LI is the cornerstone of modern physics, and all the experiments carried out so far are  consistent with it \cite{Bourgoin21,Bars19,Shao19,Bourgoin17,Flowers17,KR11},  violations of LI have been well motivated and extensively studied in the past several decades, especially from the point of view of quantum gravity   \cite{Collin04,Mattingly05,Liberati13,PT14,Wang17}.

\item    BHs in $\ae$-theory are defined in terms of universal horizons (UHs) \cite{BS11, Bhattacharyya2016, Bhattacharyya2016b}, instead of metric horizons (MHs) as that in GR. This is due to the fact that in $\ae$-theory there exist  three different species of gravitons, spin-0, spin-1, and spin-2, and each of them move in different speeds \cite{JM04}.  To avoid  the vacuum gravi-\v{C}erenkov radiation, such as cosmic rays,  each of these three species  must move with a speed that is  at least no less than the speed of light \cite{EMS05}. As a matter of fact, depending on the choice of the free coupling constants of the theory, they can be arbitrarily large, and so far no upper limits of these speeds are known \cite{KR11}.

\end{itemize}

 Clearly, now MHs are no longer one-way membranes to these particles, and they can cross MHs to escape to infinities even initially they are trapped inside them.  On the other hand, a UH is defined as the causal boundary of particles  with arbitrarily large speeds \cite{BS11}, and once they are inside it, they can never cross it to escape to   {spatial infinities} (For a recent review of UHs, see, e.g., \cite{Wang17} and references therein).

 With the above remarkable features of $\ae$-theory, it would be very interesting and  important to find  new predictions of the theory for the BH spectroscopy \cite{Berti18}, the QNMs mentioned above, which has been extensively studied in the last couple of years in terms of GWs emitted in the ringdown  phase of  binary BHs  (BBHs) (for example, see \cite{BLT21,LCV20,FBPF20,BFPF20,GST19,IGFST19,CDV19,CN20,OC20,Shaik20,Uchi20,BCCK20,Cabero20,Maselli20,Islam20} and references therein), and found that they are all consistent with GR within the error bars allowed by the observations of the 90 GW events \cite{LVK2022}.

As the first step to study QNMs in $\ae$-theory, recently some of the current authors studied spherically symmetric static spacetimes and  obtained various numerical solutions \cite{Chao2020b}, with the choice of the four free parameters  involved in the theory that satisfy all the observational constraints  \cite{OMW18} \footnote{It should be noted that all the solutions obtained previously do not satisfy these conditions, which were the main motivations for us to revisited this problem and obtained new (numerical) solutions in \cite{Chao2020b}.}.
We also derived  a set of analytic solutions in which the four-parameters  satisfy all the self-consistent and observational constraints
obtained in \cite{OMW18}, and showed that they are nothing but the (anti- or) de Sitter Schwarzschild black holes, in which the aether field are non-trivial but does not contribute to the curvatures of the spacetime. When the cosmological constant vanishes, the universal horizon locates precisely at $r = 3r_s/4$, where $r_s$ denotes the location of the  MH.

Lately, we extended our above studies to the stability of such obtained new black holes against odd parity perturbations, and found that they are stable provided that  \cite{TZZW21},
\bq
\lb{eq1.1}
c_4 = 0,
\eq
in addition to the constraint   \cite{OMW18},
\bq
\lb{eq1.2}
c_{13} \equiv c_1 + c_3 = 0,
\eq
obtained from the observations of  GW170817 \cite{GW170817} together with its gamma-ray burst GRB 170817A \cite{GRB170817A}, where $c_i$'s are the four dimensionless coupling constants of the theory.

In this paper, we continuously study  the odd-parity perturbations, but focus ourselves on the QNMs of the newly found  black holes \cite{,Chao2020b}, which, as mentioned above, satisfy all the constraints of the theory \cite{OMW18}. In doing so, several technical questions  raise. One of them is connected with the inner boundary conditions, as now the  inner boundary is the  UH, which is always inside the MH, at which the timelike Killing vector becomes spacelike. Then, the notion of ``in"- and ``out"-going waves must be define properly, before imposing the condition that there are  {\it no out-going waves} at the UH. In addition,  the field equations of perturbations 
are usually singular at the MH, and proper smoothness conditions must be imposed. In GR, the  perturbation equations are also singular at MHs, but such smoothness conditions are nicely avoided, as now the MH is the inner boundary, and the smoothness conditions are simply replaced by the inner boundary conditions.

To overcome the above issues, we first choose the ingoing Eddington-Finkelstein (EF) coordinates, so the background is smoothly across the MH, and more important, the ingoing and outgoing waves are well defined over the whole spacetime, including the region between the UH  and MH.  Then, we carefully study the structure of the field equations near the MH, and find the smoothness conditions that we must impose in order to assure that the field equations hold across the MH. Once the above issues are clarified, we first impose the purely ingoing wave conditions at the UH and purely outgoing wave conditions at the spatial infinity,  and then solve the field equations of the odd-parity perturbations  by shooting method, respectively, from the UH and the spatial infinity toward the MH. Then, at the MH we impose the smoothness conditions, which are satisfied only for particular choices of the QNM frequency $\omega$, whereby we read off  the QNMs of the odd-parity perturbations.

To show the above explicitly, in this paper we shall take the Schwarzschild black hole solution as the background, since it is also a solution of the $\ae$-theory  \cite{,Chao2020b}. The main reasons in doing are two-folds: i) First, in such a background,   the mathematics  involved will be
much simple,  from which we can construct clearly  the boundary conditions at UHs and at spatial infinities, as well as the smoothness conditions across MHs.   ii) Second,   we can understand deeply the nature of  the   instability of  this solution 
found in  \cite{TZZW21} from the point of view of QNMs, as in \cite{TZZW21} only its stability against large angular perturbations was studied.

   Specifically, the paper is organized as follows. In  Section II  we first provide a brief introduction to $\ae$-theory, and then give various definitions of horizons, including the particle, metric and universal horizons,
   which are important to the studies of QNMs of black holes, because the fact that superluminal motions exist in the theory, and the inner boundaries of black holes  now are the universal horizons. In Section III, we consider the odd-parity perturbations, and construct the  master equations, respectively, for the metric and aether field perturbations. 
    In Section IV we first work out step by step the boundary conditions for the master equation of the metric perturbations at the universal horizon as well as  at the spatial infinity. Then, we find out
the smoothness conditions across the metric horizon. Afterwards, combining the Chandrasekhar-Detweiler method \cite{Chandra1975} with the shooting method (one side  is from the UH to the MH, while the other side  is  from the spatial infinity to the MH [cf. Section IV.D]), we solve the master equation and find that only for particular choices of the mode $\omega$, can the smoothness conditions at the MH be satisfied, whereby various values of $\omega$ are found and given explicitly in Table \ref{table1} for   $l = 2, 3, 4, 10$, respectively. We also compare such a spectrum with  the corresponding one given in
GR,   and find that the differences between them   cannot be distinguished by current observations of GWs, although it is well within 
the detectability of the  next generation detectors.
In Section V,  closely following what is done  for the metric perturbations in Section IV,  we  study the aether perturbations and find various modes for the aether field with $l = 2, 10, 100$, given,
respectively, in Tables \ref{table2} and \ref{table3}, in which some unstable modes  are identified. This strongly
indicates the instability of the aether field, and is also consistent with what was found in \cite{TZZW21}. However, due to our numerical errors of the current paper, the existence of such  unstable modes  is not exclusive. In particular, in the metric perturbations presented in  Section IV,  no such instability is found. The paper is ended with Section VI, in which we derive 
our main conclusions and present some remarks on the future investigations.

Through out the paper, we shall adopt the unit system so that $c=G_N=1$, where $c$ is the speed of light while $G_N$ stands for the  Newtonian constant.   {All the Greek letters run from 0 to 3.}

\section{Einstein-aether theory}
 \renewcommand{\theequation}{2.\arabic{equation}} \setcounter{equation}{0}

In Einstein-aether theory ($\ae$-theory), the fundamental variables of the gravitational  sector are \cite{Jacobson},
\bq
\lb{2.0a}
\left(g_{\mu\nu}, u^{\mu}, \lambda\right),
\eq
where  $g_{\mu\nu}$ is the four-dimension metric  of the {spacetime}
with the  {signature} $(-, +,+,+)$,  $u^{\mu}$  {is} the aether field,
and $\lambda$ is a Lagrangian multiplier, which guarantees that the aether field  is always timelike and unity.
Then, the general action of the theory  is given  by,
\bq
\lb{2.0}
S = S_{\ae} + S_{m},
\eq
where  $S_{m}$ denotes the action of matter,  and $S_{\ae}$  the gravitational action of the $\ae$-theory, given, respectively, by
\bqn
\lb{2.1}
S_{\ae} &=& \frac{1}{16\pi G_{\ae} }\int{\sqrt{- g} \; d^4x \Big[  {\cal{L}}_{\ae}\left(g_{\mu\nu}, u^{\alpha}, c_i\right)}\nb\\
&& ~~~~~~~~~~~~ + {\cal{L}}_{\lambda}\left(g_{\mu\nu}, u^{\alpha}, \lambda\right)\Big],\nb\\
S_{m} &=& \int{\sqrt{- g} \; d^4x \Big[{\cal{L}}_{m}\left(g_{\mu\nu}, u^{\alpha}; \hat\psi\right)\Big]}.
\eqn
Here $\hat\psi$   collectively denotes the matter fields, and $g$ is the determinant of $g_{\mu\nu}$, and
\bqn
\lb{2.2}
{\cal{L}}_{\lambda}  &\equiv&  \lambda \left(g_{\alpha\beta} u^{\alpha}u^{\beta} + 1\right),\nb\\
{\cal{L}}_{\ae}  &\equiv& R(g_{\mu\nu}) - M^{\alpha\beta}_{~~~~\mu\nu}\left(D_{\alpha}u^{\mu}\right) \left(D_{\beta}u^{\nu}\right),
\eqn
where $D_{\mu}$ denotes the covariant derivative with respect to $g_{\mu\nu}$, $R$ is the Ricci scalar, and  $M^{\alpha\beta}_{~~~~\mu\nu}$ is defined as
\bqn
\lb{2.3}
M^{\alpha\beta}_{~~~~\mu\nu} \equiv c_1 g^{\alpha\beta} g_{\mu\nu} + c_2 \delta^{\alpha}_{\mu}\delta^{\beta}_{\nu} +  c_3 \delta^{\alpha}_{\nu}\delta^{\beta}_{\mu} - c_4 u^{\alpha}u^{\beta} g_{\mu\nu},\nb\\
\eqn
with $\delta_{\mu \nu}$ representing the Kronecker delta. Note that here we assume that matter fields couple not only to $g_{\mu\nu}$ but also to the aether field $u^{\mu}$. However, in order to satisfy the severe observational constraints,  such a coupling in general is assumed to be absent  \cite{Jacobson}.

The four coupling constants $c_i$'s are all dimensionless, and $G_{\ae} $ is related to  the Newtonian constant $G_{N}$ via the relation \cite{CL04},
\bq
\lb{2.3a}
G_{N} = \frac{G_{\ae}}{1 - \frac{1}{2}c_{14}},
\eq
where $c_{ij} \equiv c_i+c_j$.

The variations of the total action, respectively, with respect to $g_{\mu\nu}$,  $u^{\mu}$   and $\lambda$ yield the field equations \cite{Chao2020b},
\bqn
\lb{2.4a}
R^{\mu\nu} - \frac{1}{2} g^{\mu\nu}R - S^{\mu\nu} &=& 8\pi G_{\ae}  T^{\mu\nu},\\
\lb{2.4b}
\AE_{\mu} &=& 8\pi G_{\ae}  T_{\mu}, \\
\lb{2.4c}
g_{\alpha\beta} u^{\alpha}u^{\beta} &=& -1,
\eqn
where   $R^{\mu \nu}$ denotes the Ricci tensor, and
\bqn
\lb{2.5}
S_{\alpha\beta} &\equiv&
D_{\mu}\Big[J^{\mu}_{\;\;\;(\alpha}u_{\beta)} + J_{(\alpha\beta)}u^{\mu}-u_{(\beta}J_{\alpha)}^{\;\;\;\mu}\Big]\nb\\
&& + c_1\Big[\left(D_{\alpha}u_{\mu}\right)\left(D_{\beta}u^{\mu}\right) - \left(D_{\mu}u_{\alpha}\right)\left(D^{\mu}u_{\beta}\right)\Big]\nb\\
&& + c_4 a_{\alpha}a_{\beta}    + \lambda  u_{\alpha}u_{\beta} - \frac{1}{2}  g_{\alpha\beta} J^{\delta}_{\;\;\sigma} D_{\delta}u^{\sigma},\nb\\
\AE_{\mu} & \equiv &
D_{\alpha} J^{\alpha}_{\;\;\;\mu} + c_4 a_{\alpha} D_{\mu}u^{\alpha} + \lambda u_{\mu},\nb\\
T^{\mu\nu} &\equiv&  \frac{2}{\sqrt{-g}}\frac{\delta \left(\sqrt{-g} {\cal{L}}_{m}\right)}{\delta g_{\mu\nu}},\nb\\
T_{\mu} &\equiv& - \frac{1}{\sqrt{-g}}\frac{\delta \left(\sqrt{-g} {\cal{L}}_{m}\right)}{\delta u^{\mu}},
\eqn
with
\begin{equation}
\lb{2.6}
J^{\alpha}_{\;\;\;\mu} \equiv M^{\alpha\beta}_{~~~~\mu\nu}D_{\beta}u^{\nu}\,,\quad
a^{\mu} \equiv u^{\alpha}D_{\alpha}u^{\mu}.
\end{equation}
From Eq.(\ref{2.4b}),  we find that
\bq
\lb{2.7}
\lambda = u_{\beta}D_{\alpha}J^{\alpha\beta} + c_4 a^2 - 8\pi G_{\ae}  T_{\alpha}u^{\alpha},
\eq
where $a^{2}\equiv a_{\lambda}a^{\lambda}$.

It is easy to show that the Minkowski spacetime  is a solution of  $\ae$-theory, in which the aether is aligned along the time direction, ${u}_{\mu} = \delta^{0}_{\mu}$.
Then, the linear perturbations around the Minkowski background show that the theory in general possess three types of excitations, scalar  (spin-0), vector (spin-1) and tensor (spin-2)
modes, with their squared  speeds given by   \cite{JM04}
 \begin{eqnarray}
\label{speeds}
c_S^2 & = & \frac{c_{123}(2-c_{14})}{c_{14}(1-c_{13}) (2+c_{13} + 3c_2)}\,,\nonumber\\
c_V^2 & = & \frac{2c_1 -c_{13} (2c_1-c_{13})}{2c_{14}(1-c_{13})}\,,\nonumber\\
c_T^2 & = & \frac{1}{1-c_{13}},
\end{eqnarray}
respectively. Here $c_{ijk} \equiv c_i+c_j+c_k$.

Requiring that the theory:  (a)  be self-consistent, such as free of ghosts; and (b) satisfies all the observational constraints obtained so far,
it was found that   the parameter space of the theory is considerably restricted. In particular,    $c_{14}$, $c_2$ and $c_{13}$  are restricted to
\cite{OMW18} \footnote{The recent studies of the neutron binary systems showed that the PPN parameter $\alpha_1$ is further restricted to
$|\alpha_1| < 10^{-5}$ \cite{GHBBCYY}, which is an order of magnitude   stronger than the bounds from lunar laser ranging experiments \cite{MWT08}.
This will translate  the constraint on $c_{14}$ given by Eq.(\ref{c1234a}) to
$0 \lesssim c_{14} \lesssim  \times 2.5 \times 10^{-6}$, as one can see clearly from Eq.(3.12) given  in \cite{OMW18}.},
\bqn
\lb{c1234a}
&& 0 \lesssim c_{14} \lesssim 2.5 \times 10^{-5},\\
\lb{c1234ab}
&&  c_{14} \lesssim c_2 \lesssim  0.095,\\
\lb{c1234ac}
&& \left|c_{13}\right|  \lesssim 10^{-15}.
\eqn

Finally,  the stability of the odd-parity perturbations of BHs further requires  $c_4 = 0$   {[cf. Eq.\eqref{eq1.1}]},
provided that the condition $c_{13} = 0$ holds precisely  \cite{TZZW21}.


\subsection{Particle and Universal Horizons}

In this subsection, we shall provide a brief review on several horizons, including universal horizons, as they are important to the investigations of the QNMs of black holes
in gravitational theories with broken LI. Although they will be reviewed in the framework of the Einstein-aether theory, their generalizations to other theories   {are straightforward}.
For more details, we refer readers to \cite{Wang17}.

\subsubsection{Particle   Horizons}

In particular,  the $\ae$-theory possesses three different modes, and all of them are moving in different speeds. In general each of these different modes defines a particle horizon   \cite{Jacobson}.
These horizons are the null surfaces of the effective metrics,
\bq
\lb{EHM}
g_{\alpha \beta}^{(A)}  \equiv g_{\alpha \beta} - \left(c_{A}^2 - 1\right) u_{\alpha}u_{\beta},
\eq
where $A = S, V, T$. The null surfaces for $A = S,\;V,\;T$ are called spin-0 horizon (S0H), spin-1 horizon (S1H) and spin-2 horizon (S2H),  respectively.  These three different horizons   {will be}  referred as particle horizons for the corresponding gravitons.

In the spherically symmetric case, the static spacetimes are described by the general metric
\bqn
\lb{metric2}
ds^2 &=& -F(r) dv^2+2 B(r) dv dr+r^2 d\Omega^2,
\eqn
  {where $d \Omega^2 \equiv d\theta^2 + \sin^2\theta d\varphi^2$ } and ${x}^\mu = (v, r, \theta, \varphi)$ denote the  ingoing  EF coordinates. In this system of coordinates, the aether field takes the form
\bqn
\lb{eq2.15}
u^{\alpha} \partial_{\alpha} &=& A(r) \partial_{v}-\frac{1-F(r) A^{2}(r)}{2 B(r) A(r)} \partial_{r},
\eqn
where $A(r)$ is an arbitrary function of $r$, and  will be determined by the Einstein-aether field equations, together with the metric coefficients
$F(r)$ and $B(r)$.

Then, the out-going normal vector to the hypersurface $r = r_0$ is given by
\bqn
\lb{eq2.15a}
N_{\alpha}  \equiv \frac{\partial(r-r_0)}{\partial x^{\alpha}} = \delta^r_{\alpha},
\eqn
where $r_0$ is a constant. In terms of $N_\alpha$, the spin-0, spin-1, and spin-2 horizons are defined, respectively, by
\bqn
\lb{nullsurface}
\left. g_{\alpha \beta}^{(A)} N^{\alpha} N^{\beta} \right|_{r=r_A} = 0,
\eqn
where $r_A$ denotes the location of the horizon of the particle with spin-A.  

On the other hand,   the MH is the null surface of metric $g_{\alpha \beta}$, or a particle horizon of $g_{\alpha \beta}^{(A)} $ with $c_A  = 1$, given by
\bqn
\lb{nullsurface2}
\left. g_{\alpha \beta} N^{\alpha} N^{\beta} \right|_{r=r_{\text{MH}}} = 0.
\eqn

\subsubsection{Universal   Horizons}

To define the universal horizon, let us first note that,  if a BH is defined to be a region that traps all possible causal influences, it must be bounded by a horizon corresponding to the fastest speed.
  In theories with the broken LI,  the dispersion relation of a massive particle contains generically high-order momentum terms \cite{Wang17},
\bq
\lb{eq5.12a}
E^{2} = m^{2} +  p_{k}^2k^{2}\left(1 + \sum^{2(z-1)}_{n=1}{q_{n}\left(\frac{k}{M_{*}}\right)^{n}}\right),
\eq
from which we can see that both of the group and phase velocities
become unbounded as $k \rightarrow \infty$,
where $E$ and $k$ are the energy and momentum of the particle considered, and $p_k$ and $q_n$'s are coefficients, depending on the
species of the particle, while $M_{*}$ is the suppression energy  scale of the higher-dimensional operators. Therefore, in theories with the broken LI,
a BH should be defined to be a region that traps all possible causal influences, including particles with arbitrarily large speeds $(c_{A} \rightarrow \infty)$.

\begin{figure}[tbp]
\centering
\includegraphics[width=8cm]{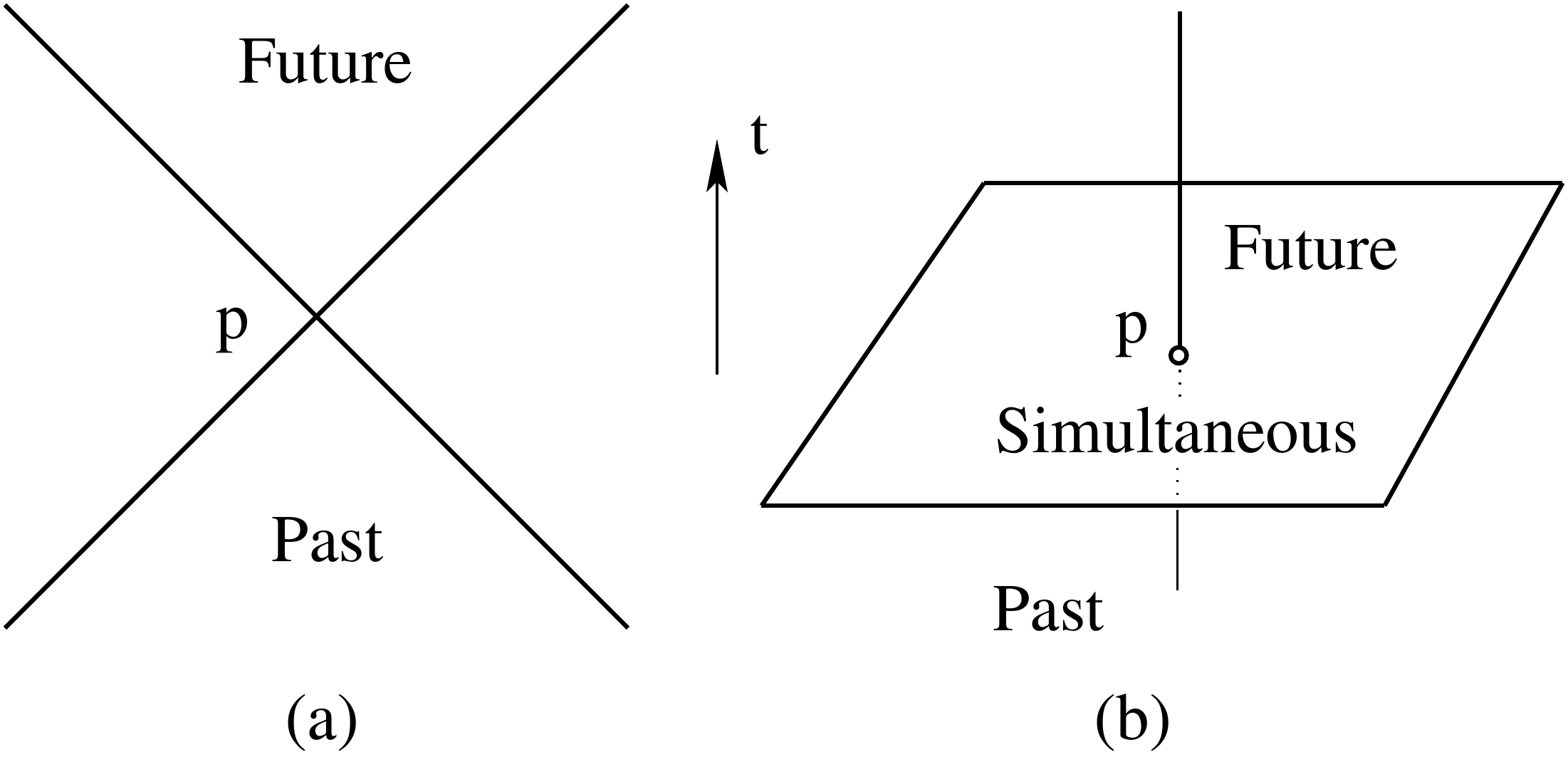}
\caption{Illustration of causal structures of spacetimes  in different theories of gravity \cite{GLLSW}:  (a) The light cone of the event $p$ in special relativity.
(b) The causal structure of the event $p$  in  Newtonian theory. }
\label{fig1}
\end{figure}

Does such a region exist? To answer this question, let us first note that    the causal
structure of spacetimes in such theories is quite different from that given in GR,  where the light cone at a given point $p$ plays a fundamental
role in determining the causal relationship of $p$ to other events \cite{GLLSW}.  In a ultraviolet (UV) complete theory, the above dispersion relationship is expected even in the
gravitational sector \footnote{One of such examples is  the  healthy extension  \cite{BPSa,BPSb} of Ho\v{r}ava gravity \cite{Horava,Wang17}, a possible UV extension of
the khronometric theory.}.  In such theories,  the causal  structure is dramatically changed.
For example, in  Newtonian theory,  time is absolute and the speeds of signals are not limited. Then, the
causal structure of  a given point $p$ is uniquely determined by the time difference, $\Delta{t} \equiv t_{p} - t_{q}$, between the two events [cf. Fig. \ref{fig1}].
 In particular, if $\Delta{t} > 0$, the event $q$ is to the past of $p$; if $\Delta{t} < 0$, it  is to the future; and if $\Delta{t} = 0$, the two events are
simultaneous.

\begin{figure}[tbp]
\centering
\includegraphics[width=1\columnwidth]{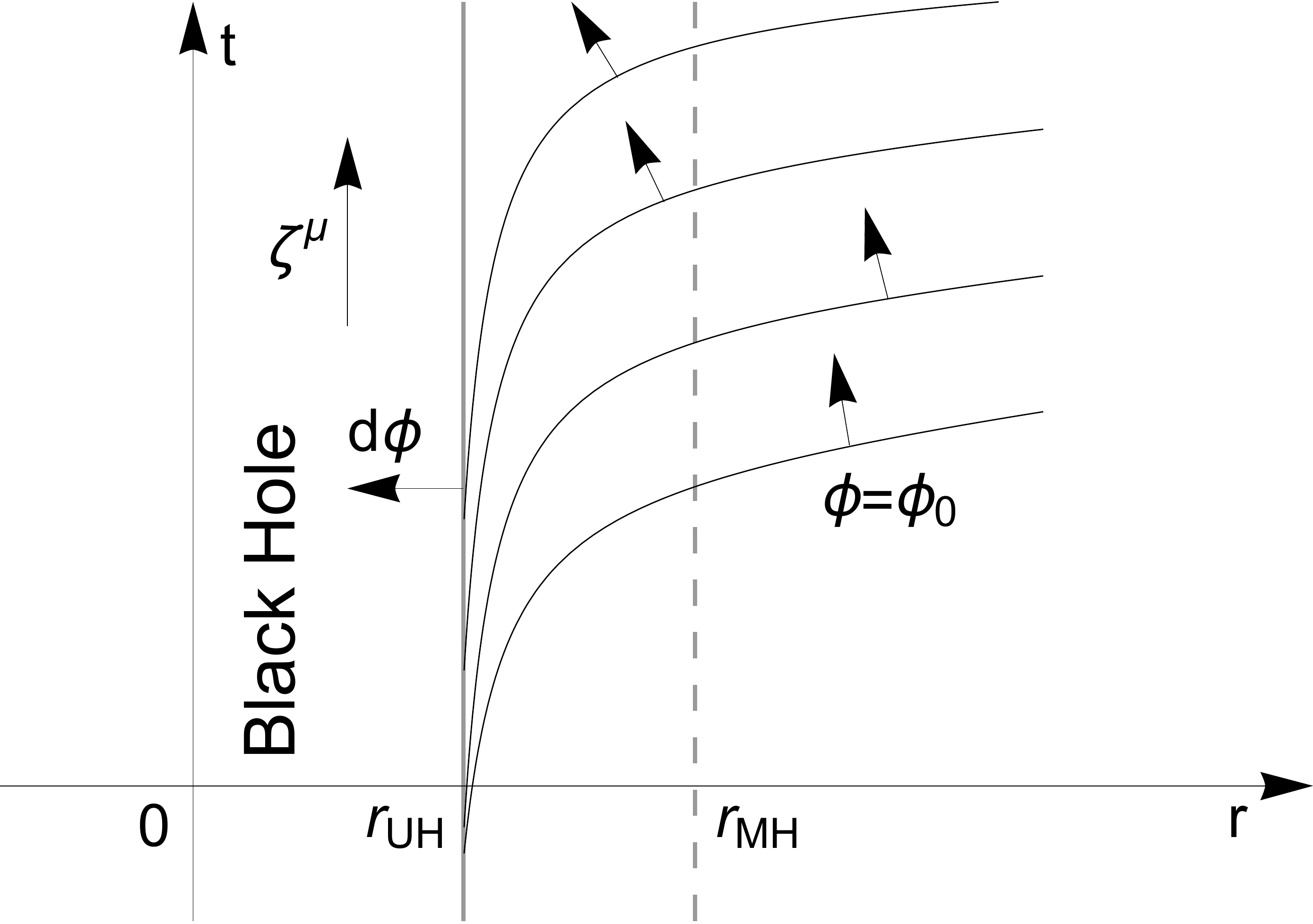}
\caption{Illustration of the bending of the $\phi$ = constant surfaces, and the existence of the UH  in a spherically symmetric static spacetime,
where $\phi$  denotes the  globally timelike scalar  field, and $t$  is the Painlev\'{e}-Gullstrand-like coordinates, which covers the whole spacetime   \cite{LSW16}. Particles
move always along the increasing direction of $\phi$.
The Killing vector $\zeta^{\mu} = \delta^{\mu}_{v}$ always points upward at each point of the plane. The vertical dashed  line is
 the location of the metric (Killing) horizon, $r=r_{\text{MH}}$. The UH, denoted by the vertical solid  line, is  located at $r = r_{\text{UH}}$, which is always inside the MH.}
\label{fig2}
\end{figure}

 In  theories with   {broken LI},  a similar situation occurs. Thus, to consider the  causal
structure of spacetimes in such theories,  a globally time-like   ``coordinate"  needs to be introduced \cite{BS11} (See also \cite{BJS11}).
 In particular, for a given spacetime,  let us  first introduce a globally timelike scalar field  $\hat\phi$  \cite{LACW}, so that
 \bq
 \lb{eq5.12aa}
 g^{\alpha\beta}\hat\phi_{,\alpha}\hat\phi_{,\beta} > 0,
 \eq
 over the whole spacetime.  It is interesting to note that,
in spherically symmetric  spacetimes,  the timelike aether field $u_{\mu}$ is hypersurface-orthogonal \cite{Jacobson10,Wang17},
and   {can always be} expressed as the gradient of a scalar field  \cite{Wald94},
\bq
\lb{eq2.13}
u_{\mu} = \frac{\phi_{,\mu}}{\sqrt{-\phi_{,\alpha}\phi^{,\alpha}}}.
\eq
Therefore, now a natural choice is to identify $\hat\phi$ with   $\phi$, $\hat\phi = \phi$.
Then, similar to
Newtonian theory, this field defines  globally an absolute time direction: {\it all particles are assumed to move along the increasing direction of the timelike scalar field $\phi$}. It is clear with this definition of the time direction,  the causality now is well-defined, quite similar to  Newtonian theory [cf. Fig. \ref{fig2}. See also  \cite{Enrico2013}].  With such a definition, one can see that, for a given spacetime there may exist a surface $r = r_{\text{UH}}$, at which the  aether field $u_{\mu}$ is orthogonal to the timelike Killing vector  \cite{LGSW15}
 $\zeta\; (\equiv \partial_{v})$,
 \bq
\lb{eq5.12}
\left.\zeta \cdot u\right|_{r = r_{\text{UH}}}   = -\left. \frac{1}{2A}\left(1 + J\right)\right|_{r = r_{\text{UH}}} = 0,
\eq
where $J \equiv FA^2$  \cite{Chao2020b}.  Given that all particles move along the increasing direction of the  aether field,   it is clear that a particle must cross this surface and move inward,
once it arrives at it, no matter how large  its speed  is \footnote{ Particles even with infinitely large speeds   would just move on these
boundaries and  cannot escape to infinity.}. This is a one-way membrane, and particles even with infinitely large speeds cannot escape from it, once they are
inside it. So, it acts as an absolute horizon to all particles (with any speed),  which is often called the UH  \cite{BS11,BJS11,Wang17}.

The extension of the definition of UHs to dynamical spacetimes is given in \cite{BMSW18}, in which it was shown explicitly that such dynamical UHs  can be formed from gravitational
collapse of realistic matter fields.

\subsection{The de Sitter-Schwarzschild Black Hole Solutions}

 When $c_{13}=c_{14}=0$, there exists a particular class of analytic solutions of the background Einstein-aether field equations \cite{Chao2020b},
\bqn
\lb{anasolu1}
F(r) &=& F_2\left(1 - \frac{2m}{r}\right) + \frac{\Lambda}{3}r^2,\nb\\
B  &=& \sqrt{F_2}, \quad \Lambda \equiv \frac{9}{8}c_2  w_1^2w_2^2,\nb\\
A(r) &=& - \frac{w_2}{2F} \Bigg[\left(\frac{1}{r^2} + w_1 r\right) \nb\\
&&   \left. \pm   \sqrt{\frac{4F}{w_2^2} +\left(\frac{1}{r^2} + w_1 r\right)^2}\right],
\eqn
where   $F_{2}, \; m, \; w_{1}$ and $w_{2}$ are four integration constants. Using the gauge freedom of the EF coordinates, without loss of the generality, we can always
set $F_2 = 1$. Then, the spacetime is precisely the anti-de Sitter or de Sitter Schwarzschild black hole solutions, depending on the signs of the coupling constant
$c_2$. However, the constraints (\ref{c1234a}) and (\ref{c1234ab}) tell us that $c_{2} \ge 0$, so it must be the de Sitter Schwarzschild solution, where
$r_s = 2m$ is the Schwarzschild radius. In this paper, we shall consider only the case $w_1 = 0$, so that the spacetime is described precisely by the
Schwarzschild solution. Then,  as $r \rightarrow \infty$, we find that
\bq
\lb{anasolu1aa}
\lim_{r \to \infty} A(r) = \mp 1, \; (F_2 = 1, \; w_1 = 0).
\eq
Thus, we shall choose the ``-" sign   {for the expression of $A(r)$ in Eq.\eqref{anasolu1}}, so that    $\lim_{r \to \infty} A(r) = + 1$, which is also consistent
 with the asymptotical-flatness conditions adopted in \cite{Chao2020b}.
  Then, for a UH to exist, we must   set $w_2 = 3\sqrt{3} r_s^2/8$ \cite{LGSW15}, for which the UH is located at
\bqn
\lb{rUH}
r_{\text{UH}} &=& \frac{3}{4} r_{s}.
\eqn
Note that in the current case we have
\bq
\lb{css}
c_S = \infty, \quad c_T = c_V = 1, \; (c_{13} = c_{14} = 0),
\eq
as can be seen from Eq.~(\ref{speeds}), which implies that now the S0H coincides with the UH, while the S1H and S2H coincide with the MH.

It is interesting to note that in this particular case, the aether field has no influence on the spacetime geometry. However,
the aether field is non-trivial, and quite different from the one given in the Minkowski spacetime.  This will in turn affect the perturbation equations of the Einstein-aether gravity, as to be seen clearly in the next section.


\section{Odd-Parity Perturbations and Field Equations}
\renewcommand{\theequation}{3.\arabic{equation}} \setcounter{equation}{0}

To study the  linear perturbations of the spherically symmetric spacetimes in the framework of $\ae$-theory, let us first  introduce the  more familiar Schwarzschild coordinate $t$ via
the relation,
\bqn
\lb{vandt}
t = v -  f(r), \quad df \equiv \frac{B}{F} dr,
\eqn
from which we find that the background metric (\ref{metric2})   {and background aether field \eqref{eq2.15} take} the form \cite{TZZW21},
\bqn
\lb{metric3}
d {\tilde{s}}^2 &=& -F(r) dt^2+ \frac{B^2(r)}{F(r)} dr^2 +r^2 d\Omega^2,\\
\lb{aether3}
{\tilde {u}}^\alpha \partial_{\alpha} &=& \frac{FA^2 + 1}{2AF}\partial_{t} +  \frac{FA^2 - 1}{2AB}\partial_{r}.
\eqn
Clearly, this metric becomes singular at the MH, $F(r_{s}) = 0$, so that the coordinates cannot cover   the whole region $r \in (0, \infty)$.
 However, now the inner boundary is not at  the MH, but rather at the UH,  which is always inside the MH. Therefore, to impose the  boundary conditions
 at $r = r_{\text{UH}}$ and $r =  \infty$ simultaneously, we shall work out the linear perturbations  in the   EF  coordinates, in which the metric takes the form of Eq.(\ref{metric2}).  On the other hand, setting
 \bqn
\lb{vandtB}
u \equiv  t - f(r),
\eqn
we find that  the metric (\ref{metric3}) takes the form
\bqn
\lb{metric4}
d{\bar s}^2   &=& -F(r) du^2 - 2 B(r)dudr  +r^2 d\Omega^2,
\eqn
which is the  metric written in terms of  the out-going EF  coordinate $u$.
Although the roles of $t$ and $r$ will be exchanged across a MH, the ``in-going" and ``out-going" coordinates $v$ and $u$ remain the same in the whole spacetime. So, in the rest of the paper we always refer the ``in-going" and ``out-going" in terms of the coordinates $v$ and $u$.

Once the above issues are clarified, now let us consider the perturbations of the background solutions given by Eq.(\ref{anasolu1}) with
\bq
\lb{eq3.0}
F_2 = 1, \quad w_1 = 0, \quad w_2 = \frac{3\sqrt{3} r_s^2}{8}.
\eq
For such choices, the background spacetime is precisely the Schwarzschild vacuum solution,
\bq
\lb{eq3.0a}
F = 1 - \frac{r_s}{r}, \quad B = 1,
\eq
as shown in the last section.

Denoting  the background metric and aether field by $\bar g_{\mu \nu}$ and $\bar u_\mu$, respectively,  the total metric and aether field are given by,
\bqn
\lb{gab}
g_{\mu \nu} &=& \bar g_{\mu \nu}+ \epsilon h_{\mu \nu}, \quad
u_{\mu} = \bar u_{\mu} + \epsilon w_{\mu},
\eqn
where $\epsilon$ is a book-marker, and we shall expand the perturbations only to its first order.  Later, we can safely set it to one.
Then,  working in the in-going EF coordinates $(v, r, \theta, \phi)$, and with only the
the odd-parity part, we find that   the linear perturbations
$h_{\mu\nu}$ and $w_{\mu}$ can be cast in the forms \cite{Thomp2017, TZZW21},
\begin{widetext}
	\bqn
	\lb{hab}
	h_{\mu \nu} &=&  \sum_{l=0}^{\infty} \sum_{m=-l}^{l}
	\begin{pmatrix}
		0	& 0 &  C_{lm}\csc \theta \partial_\varphi & -C_{lm} \sin \theta \partial_\theta\\
		0	& 0 &  J_{lm}\csc \theta \partial_\varphi & -J_{lm} \sin \theta \partial_\theta\\
		sym	& sym &  G_{lm}\csc \theta \big(\cot \theta \partial_\varphi-\partial_\theta \partial_\varphi \big) &  sym \\
		sym	& sym &  \frac{1}{2} G_{lm}\big(\sin \theta \partial^2_\theta - \cos \theta \partial_\theta- \csc \theta \partial^2_\varphi\big) &   - G_{lm} \sin \theta \big(\cot \theta \partial_\phi-  \partial_\theta \partial_\varphi \big)
	\end{pmatrix}
	Y_{l m}(\theta, \varphi), 	\nb\\
\eqn
\end{widetext}
and
\bqn
\lb{wa}
	w_\mu &=& \sum_{l=0}^{\infty} \sum_{m=-l}^{l}
	\begin{pmatrix}
		0	\\
		0	\\
		a_{lm} \csc \theta  \partial_\varphi 	\\
		-a_{lm} \sin \theta  \partial_\theta
	\end{pmatrix}
Y_{l m}(\theta, \varphi),
\eqn
where  $Y_{lm}(\theta, \varphi)$ stands for the spherical harmonics, and $C_{lm}, \; G_{lm},\; J_{lm}$ and $a_{lm}$ are functions of $v$ and $r$ only \footnote{
It must not be confused with  the function {$J_{lm}$} introduced  here  and the tensor {$J_{\alpha \beta}$} appearing in Eq.\eqref{2.5}.}.   Note that when calculating the field equations   we will set  $m=0$ in the above expressions so that $\partial_\varphi Y_{lm}(\theta, \varphi)=0$,
as now the background has the spherical symmetry, and the corresponding linear perturbations do not depend on $m$   \cite{Regge57,Thomp2017} \footnote{Notice that, since we are using the EF coordinate and following some different conventions, the terms $\{C_{lm}, J_{lm}, G_{lm}\}$ in \eqref{hab} are not necessarily equal to their counterparts given in \cite{Thomp2017}.}.

\subsection{Gauge Transformations}

For later convenience, we first investigate the infinitesimal gauge transformations (Recall that we only consider the odd-parity perturbations and $m=0$)
\bqn
\lb{gauge1}
x^\alpha \to x^{\prime \alpha} =x^\alpha+  \epsilon \xi^{\alpha},
\eqn
where
\bqn
\lb{xi}
\xi^\alpha &=& - \frac{\csc \theta \partial_\theta Y_{lm}(\theta, \phi)}{r^2}  \left\{0,0,0,1 \right\} \xi,~~~~
\eqn
with $\xi$ being a function of $v$ and $r$.
Under the transformation of Eq.(\ref{gauge1}), we have  
\bqn
\lb{deltau}
&& \Delta w_\mu \equiv \left(w_\mu\right)_{new}-\left(w_\mu\right)_{old} =- {\cal{L}}_{\xi} {\bar u}_\mu, \nb\\
&& \Delta h_{\mu\nu} \equiv \left(h_{\mu\nu}\right)_{new}-\left(h_{\mu\nu}\right)_{old} =- {\cal{L}}_{\xi} {\bar g}_{\mu\nu},
\eqn
where $\cal{L}$ stands for the Lie derivative \cite{Invb}.  From Eq.(\ref{deltau}) we find \cite{TZZW21}
\bqn
\lb{gauge2}
&& \Delta C_{lm} \equiv \left( C_{lm}\right)_{old}-\left( C_{lm}\right)_{new} = {\dot \xi}, \nb\\
&& \Delta G_{lm} \equiv \left( G_{lm}\right)_{old}-\left( G_{lm}\right)_{new} = -2 \xi, \nb\\
&& \Delta J_{lm} \equiv \left( J_{lm}\right)_{old}-\left( J_{lm}\right)_{new}  = \xi' - \frac{2}{r} \xi, \nb\\
&& \Delta a_{lm} \equiv \left( a_{lm}\right)_{old}-\left( a_{lm}\right)_{new} = 0,
\eqn
where a prime and a dot stand for the derivatives with respect to $r$ and $v$, respectively.
 With the above gauge transformations,  we can construct the gauge-invariant quantities, and due to the presence of the aether field,
 three such independent quantities can be constructed, in contrast to the relativistic case, in which only two such quantities can be constructed.
 These three gauge invariants  can be  defined as
\bqn
\lb{gauge4}
{\cal{X}}_{lm} (v, r) &\equiv& {C}_{lm} +  \frac{1}{2} {\dot G}_{lm}, \nb\\
{\cal{Y}}_{lm} (v, r) &\equiv&  \frac{2}{r} C_{lm} - C_{lm}' +  {\dot J}_{lm}, \nb\\
{\cal{Z}}_{lm} (v, r) &\equiv& a_{lm}.
\eqn
Of course, any combination of these quantities is also gauge-invariant.
By properly   fixing the gauge, the resultant field equations will be simplified considerably.  In the current case, we find that
 one of the most convenient gauge choices  is
 \bq
 \lb{gauge4aa}
 C_{lm} =0,
 \eq
 from which
we find
\bqn
\lb{gauge3}
\xi(v, r) &=& \int \left[C_{lm}(v, r)\right]_{old} dv + {\hat C}_1 (r),
\eqn
where $ {\hat C}_1$ is an arbitrary function of $r$. Therefore, with this choice of gauge, we still have the gauge residual
\bq
\lb{gauge4bb}
{\hat \xi}(v, r) = {\hat C}_1 (r).
\eq
We shall come back to this point later, and fix the gauge residual completely by properly imposing an additional gauge condition.

\vspace{0.5cm}

\subsection{Linearized Field Equations}

When the spacetimes are in the vacuum, we have $T_{\mu\nu} = 0, \; T_{\mu} = 0$, and then  the field equations (\ref{2.4a}) and (\ref{2.4b}) reduce to
\bqn
\lb{fieldeqn}
E_{\mu \nu} \equiv G_{\mu \nu}-S_{\mu \nu} = 0, \\
\lb{fieldeqnb}
\AE^\mu =0,
\eqn
where $G_{\mu \nu} \equiv R_{\mu \nu}-g_{\mu \nu} R/2$.
To the first-order of $\epsilon$ and with the gauge (\ref{gauge4aa}), we find that there are only four non-trivial equations, which are given by
\bq
\lb{LQs}
E_{\phi t}=E_{\phi r}=E_{\phi \theta}=\AE^{\phi}=0,
\eq
 and can be cast respectively in the forms given by Eqs.(\ref{fieldeqn1}) - (\ref{fieldeqn4}) in Appendix A.

It is interesting to note that the coefficients of these differential equations, i.e., $\rho_{abc}$ given by Eqs.(\ref{rho101}) - (\ref{rho401}), are functions of  $r$, $c_1$ and $l$ only, and some of them contain no $c_1$.  
More importantly, from Eq.~\eqref{fieldeqn4} we can see clearly that $a_{lm}$ has been decoupled from $J_{lm}$ and $G_{lm}$.

\subsection{Master Equations}

Since there are only three independent unknowns in Eqs.~\eqref{fieldeqn1}-\eqref{fieldeqn4}, $G_{lm}, \; J_{lm}$ and $a_{lm}$, one of the above four equations must not be independent, although   this non-dependent equation can still be used to simplify the derivations of  the master equation. To simplify the following discussions, using the rescaling symmetry of the background \cite{Chao2020b}, we first set  $r_{s}=1$. Then, we have
\bq
\lb{sch_f}
f(r) \equiv \int{\frac{B}{F} dr} = r + \ln\left|r - 1\right|.
\eq
 {To find the master equations for the gauge invariants introduced in Eq.(\ref{gauge4}), we first eliminate the ${\ddot a}_{lm}$ terms
from  Eqs.(\ref{fieldeqn1}) and (\ref{fieldeqn2}). Then, combining this resultant equation with Eq.(\ref{fieldeqn3}) we find that the terms proportional to  ${\dot G}_{lm}'$ and ${\dot G}_{lm}$ can be made disappear. Combining
this resultant equation with Eq.(\ref{fieldeqn1}) and the first-order $v$-derivative of Eq.(\ref{fieldeqn2}), we obtain an equation that contains 
  the ${\dddot a}_{lm}$ and ${a}_{lm}''$ terms. Finally, combining this new resultant equation   with Eq.(\ref{fieldeqn4}) and its $v$ derivative, we  arrive
  at the master equation}
%
%
\begin{widetext}
\bqn
\lb{master1}
&& \Bigg\{ \frac{(r-1)^2}{r^2} \frac{\partial^2}{\partial r^2} +\frac{(r-1) (2 r-1)}{r^3} \frac{\partial}{\partial r}   
+2 \left( 1-\frac{1}{r} \right) \frac{\partial^2}{\partial r \partial v} + \frac{2 (r-1)}{r^2} \frac{\partial}{\partial v}   - \frac{(r-1) [ l (l+1) r-4]}{r^4}\Bigg \}{\cal{Y}}_{lm} = 0, ~~~~
\eqn
where  ${\cal{Y}}_{lm} = \dot{J}_{lm}$. 
Setting
\bqn
\lb{calY1}
{\cal{Y}}_{lm} (v,r) &=& e^{-i \omega v} {\bar {\cal{Y}}}_{lm} (r),
\eqn
where $\omega$ is the QNM frequency to be calculated below, \eqref{master1} reduces to
\bqn
\lb{master2}
&& \Bigg\{\frac{(r-1)^2}{r^2} \frac{\partial^2}{\partial r^2} +\left[\frac{(r-1) (2 r-1)}{r^3}
 -2i\omega \left( 1-\frac{1}{r} \right)\right] \frac{\partial}{\partial r} - \frac{2i\omega (r-1)}{r^2}     - \frac{(r-1) [ l (l+1) r-4]}{r^4}\Bigg \}\bar{\cal{Y}}_{lm} = 0. ~~~~
\eqn
\end{widetext}
To eliminate the first-order derivative term of the above equation,   {following the general prescription presented in Appendix B,  we  introduce the function $\Psi(r)$ via the relation,}
 \bqn
\lb{Psi}
\Psi(r) & \equiv & e^{-i\omega f(r)} r {\bar {\cal{Y}}}_{lm}(r),
\eqn
 so that Eq.~(\ref{master2}) can be cast in the simple form,
\bqn
\lb{master2a}
\left[\frac{d^2}{d x^2} + \left(\omega^2 -V_g \right) \right] \Psi =0,
\eqn
where $ x \equiv f(r)$  given by Eq.(\ref{sch_f}), and
\bqn
\lb{Vg}
V_g &\equiv&   \frac{(r-1) [l (l+1) r-3]}{r^4}.
\eqn
  {Note that, the range of the new variable {$x \in (-\infty, +\infty)$}  covers only  the region {$r\in (1, \infty)$}. Thus, as will be seen later, we have to look for another variable to cover the whole
 region $r \in \left(r_{UH}, \infty\right)$ where $r_{UH} = 3/4$. However, this equation is very useful when we study the asymptotical behavior of the perturbations at the spatial infinity $ r \rightarrow \infty$, 
 as can be seen from the discussions to be presented below  in Section IV.A.}  It is interesting to note that $V_g$ is identical to the effective potential given in the Regge-Wheeler equation in GR \cite{Kono2011}. However, due to the
different locations of the inner boundaries, in general different BH spectroscopies are expected.

On the other hand, solving for ${\dot G}_{lm}$ and $a_{lm}''$ from  Eqs.(\ref{fieldeqn1}) and (\ref{fieldeqn4}) we find
\bqn
\lb{equ7b}
  {\cal{X}}_{lm}  &=& -\frac{1}{(l+2)(l-1)} \nb\\
  && \times \left[2(r-1)+r(r-1) \frac{\partial }{\partial r} +r^2 \frac{\partial }{\partial v} \right] {\cal{Y}}_{lm},~~~~~~~
\eqn
where  Eq.\eqref{gauge4} was used in writing down the above expression.  Then,   from Eqs.\eqref{gauge4} and (\ref{gauge4aa}) we find
 \bqn
\lb{equ7c}
 G_{lm}(v, r)  &=& 2 \int{{\cal{X}}(v, r) dv} + \hat{G}_{lm}(r),
 \eqn
 where $\hat{G}_{lm}(r)$ is an arbitrary function of $r$ only. However, using the gauge residual (\ref{gauge4bb}), we can always set it to zero, by choosing
 \bq
 \lb{equ7d}
 \hat{C}(r) = -\frac{1}{2}\hat{G}_{lm}(r).
 \eq
 Clearly, with this choice the remaining gauge freedom is completely fixed.

In addition,  following Appendix B  we find that Eq.(\ref{fieldeqn4}) can be written in the form
\bqn
\lb{master4}
\left[ \frac{(r-1)^2}{r^2}  \frac{d^2}{d r^2} + \frac{r-1}{r^3}  \frac{d}{d r} - V_{\text{eff}}(r) \right ] \psi &=& 0, ~~~~
\eqn
where
\begin{widetext}
\bqn
\lb{psi}
\psi(r) &\equiv& \sqrt{ \frac{r}{r-1} } \exp \left\{ \int \frac{54(1-r)+  r^2\left[128 (r-1) r+3 i \sqrt{768 (r-1) r^3+81} \omega_{\ae} \right] } {(3-4 r)^2 (r-1) r [8 r (2 r+1)+3]} dr \right\} {\hat {\cal{Z}}}_{lm}(r), \\
\lb{Veff}
V_{\text{eff}} &\equiv& \frac{1}{4 \left(256 r^8-256 r^7+27 r^4\right)} \Big[1024 l (l+1) r^6-1024 \left(2 l^2+2 l+1\right) r^5  +256 \left(4 l^2+4 l+9\right) r^4\nb\\
&&~~~~~~~~~~~~~~~~~~~~~~~~~~~~~~~~~~  -1280 r^3+2160 r^2-4212 r+2079 \Big],
\eqn
\end{widetext}
with
\bqn
\lb{almvr}
a_{lm}(v, r) =  e^{-i \omega_{\ae} t} {\hat {\cal{Z}}}_{lm}(r).
\eqn
Here $t$ is related to $v$ through Eqs.(\ref{vandt}) and (\ref{sch_f}).   {Notice that, a subscript $\ae$ is added to $\omega$ here to allow it to be different from
 the QNM spectra $\omega$ obtained from Eq.(\ref{master2a}).

It is remarkable to note that $a_{lm}$ depends on $\omega_{\ae}$ only through Eq.(\ref{psi}), while  Eq.(\ref{master4}) shows that $\psi$ itself does not depend on
 $\omega_{\ae}$, although its boundary conditions will depend on $\omega_{\ae}$ through $a_{lm}$. We shall come back to this point in the next section.

\section{QNMs of black holes in Einstein-aether Theory}
 \renewcommand{\theequation}{4.\arabic{equation}} \setcounter{equation}{0}

From the last section, we can see that solving the linearized Einstein-aether field equations for the odd-parity perturbations now reduces to solving the master equations (\ref{master2a}) and (\ref{master4}) with proper boundary conditions.  As in GR, they have solutions only for some particular choices of $\omega$ and $\omega_{\ae}$, which form the spectra of the QNMs of the corresponding BH.

In contrast to GR, due to the existence of superluminal propagations caused by the LI violation,  the location of the inner boundary conditions now is moved from MHs to UHs, which are always inside the MHs.
There are at least two aspects that make us to believe that  QNMs in $\ae$-theory in general  deviate from those of GR:
\begin{itemize}

 \item The different locations of the inner boundaries. It is true that in both theories it is all required that only ``in-going" waves be allowed at the inner boundaries. However, when these conditions are imposed at  UHs (as required by $\ae$-theory), it is expected that at the corresponding  MHs both ``in-going" and ``out-going" waves exist (which is quite different from that of GR), because of the existence of the superluminal propagations.

\item The different dynamical differential equations. Despite the fact that the background in both theories is the same and all described by the Schwarzschild BH, the linearized perturbation equations are different, and in general it is expected that these also alternate the BH QNMs.
\end{itemize}

With the above in mind, let us turn to consider the initial conditions respectively at the UH ($r = r_{\text{UH}} =3/4$) and   spatial infinity ($r \to +\infty$).
Once these are done, we turn ourselves to consider the smoothness conditions at the MH. 

 \subsection{Boundary Conditions at the Spatial Infinities}

As mentioned above, the boundary conditions at the spatial infinity are that only ``out-going" waves exist. To see the implication of these conditions on the gauge-invariant quantities ${\cal{Y}}_{lm}$ and $a_{lm}$, let us first note that
 as   $r \to +\infty$, to the leading order, we have $f(r) \simeq r, \; V_g(r) \rightarrow 0$, and $v \simeq t + r$ and $u \simeq t - r$. Then,  Eq.\eqref{master2a} has the general solutions,
\bq
\lb{Asmptotic1}
\Psi |_{r\to \infty}  = a_+ e^{ i \omega x} + a_- e^{ - i \omega x},
\eq
where
$a_{\pm}$ are two integration constants.
 Inserting the two branches of \eqref{Asmptotic1} into \eqref{Psi}, and together with Eq.~\eqref{calY1}, we find that
\bqn
\lb{BC1}
\left. {\cal{Y}}_{lm}(v,r) \right|_{r \to +\infty} & \propto & r a_{\pm}  e^{-i\omega(t \mp x)}.
\eqn
Therefore, to have only out-going waves, we must set $a_- = 0$.  As a result, when $r$ is very large, we expect $\Psi(r)$ to take the form,
\bqn
\lb{expand1}
\Psi &=&  e^{ i \omega x} \sum_{n=0}^\infty \frac{a_n}{r^n},
\eqn
where $a_n$'s are constants to be determined with the Frobenius-like method \cite{Butkov}. In particular, inserting
the above expression into Eq.(\ref{master2a}), we find the following  recursion relation,
\bqn
\lb{eq4.7}
&& 2i n \omega a_n - \Big[(n-1)(n+2i \omega) - l^2 - i \Big] a_{n-1}\nb\\
&& - \Big(l^2 + l - 2n^2 + 5n + 1\Big) a_{n-2} + (n-4) n a_{n-3} = 0, \nb\\ 
\eqn
from which we can write all $a_n$'s ($n \ge 1$) in terms of $a_0$   {with $a_n \propto a_0$}. 
As will be seen later, the amplitude of $\Psi(r)$ has no contributions to our main results, Therefore, without loss of the generality, we can always set
$a_0 = 1$.

\subsection{Boundary Conditions at Universal Horizons}

On the other hand, as  $r \to r_{\text{UH}} \; (=3/4)$, we note that Eq.~\eqref{master2} can be written as
\bqn
\lb{master2b}
&& \left[{\tilde W}_g(r) \frac{d^2}{d y^2} + \left(\omega^2 -{\tilde V}_g \right) \right] {\tilde \Psi} =0,\\
\lb{master2ba}
&& {\tilde \Psi} \equiv \sqrt{\frac{r(1-r)}{{\tilde p}}}  e^{-i\omega f} {\bar{\cal{Y}}}_{lm},   \quad {\tilde W}_g  \equiv  \frac{\eta_1}{{\tilde p}^2},
\eqn
where
\begin{widetext}
\bqn
\lb{rast2}
\frac{dr}{dy} & \equiv &   {\tilde p} (r) \equiv  \frac{\sqrt{l^2+l-4} + 12+l (l+1) (4 r-3)-16 r}{3 \sqrt{l^2+l-4}+12+l (l+1) (4 r-3)-16 r},\nb\\
{\tilde V}_g  &\equiv &   -\left\{ \frac{{\tilde W}_g}{4} \left[2 {\tilde p} \frac{d \left({\tilde p}'-{\tilde p} \eta_2/\eta_1 \right)}{dr}  -\left({\tilde p}'- \frac{ {\tilde p} \eta_2}{\eta_1} \right)^2 \right] + \eta_3 \right\},\nb\\
\eta_1 &\equiv& \frac{(r-1)^2}{r^2}, \quad
\eta_2 \equiv  \frac{(r-1) (2 r-1)}{r^3} , \quad
\eta_3 \equiv -\frac{(r-1) \left(l^2 r+l r-4\right)}{r^4}.
\eqn
\end{widetext}
Thus, as $r \to r_{\text{UH}}$, we have
\bqn
\lb{asymp3b}
  \frac{d^2 {\tilde \Psi}}{d y^2} +\omega^2  {\tilde \Psi} &=& 0,
\eqn
which leads to
\bqn
\lb{Asmptotic2}
{\tilde \Psi}|_{r\to  r_{\text{UH}}} &=& b_+ e^{+ i \omega  y }  + b_- e^{- i \omega  y },
\eqn
with $b_\pm$ being arbitrary constants.
Combining Eqs. \eqref{Asmptotic2}, (\ref{master2ba}) with Eq.~\eqref{calY1}, we find that
\bqn
\lb{BC2}
\left. {\cal{Y}}_{lm}(t,r) \right|_{r \to  r_{\text{UH}}} & \propto &  b_{\pm} e^{ -i \omega \left( t \mp  y \right)} \sqrt{\frac{{\tilde p}}{r(1-r)}} \nb\\
& \propto &  b_{\pm} e^{ -i \omega \left(t \mp  r \right)} .
\eqn
Therefore, to satisfy the pure ingoing-wave conditions, we must set {$b_+ = 0$}.  As a result, when $r \to  r_{\text{UH}}$, we expect ${\tilde \Psi}(r)$ to take the form,
\bqn
\lb{expand3}
{\tilde \Psi} &=&  e^{- i \omega y} \sum_{n=0}^\infty b_n \left(r- \frac{3}{4} \right)^n.
\eqn
Similar to $a_n$'s, inserting the above expression into Eq.(\ref{master2b}) we shall find a recursion relation for $b_n$'s, and from which we can
write all $b_{n \ge1}$'s in terms of $b_0$   {with $b_n \propto b_0$}.  Again, as will be seen, the choice of $b_0$ is actually irrelevant to our solutions of QNMs, so without loss of the generality, we can also set it to one, $b_0 = 1$.

\subsection{Smoothness Conditions across Metric Horizons}

 In principle, once the two boundary conditions are given, we can solve the two master equations (\ref{master2a}) and (\ref{master4})
to find out   {the spectrum of $\omega$ ($\omega_{\ae}$)} for any given $l$. However, since both of them are  singular at the
MH, we have to solve  them first in the regions $r \in (r_{UH}, 1-\epsilon)$ and $r \in (1+\epsilon, \infty)$ separately, and then match the solutions together
across the MH, where $0< \epsilon \ll 1$ [Notice that $\epsilon$ used here must  not confused with the ones appearing in Eq.\eqref{gab}].

To find out the proper matching conditions in the neighborhood of $r=1$, Eq.(\ref{master2a}) cannot be used, as it is valid only for $r \in (1, \infty)$. To overcome this problem, 
let us first replace $ {\bar {\cal{Y}}}_{lm}$ appearing in Eq.~\eqref{master2}  by
\bq
\lb{4.21a}
 {\hat {\cal{Y}}}_{lm} \equiv e^{-i\omega f} {\bar {\cal{Y}}}_{lm}, 
 \eq
 so that it  reduces to
\begin{widetext}
\bqn
\lb{master2c}
\Bigg\{\frac{(r-1)^2}{r^2} \frac{d^2}{d r^2} + \frac{(r-1) (2 r-1)}{r^3} \frac{d}{d r} + \left[ \omega ^2-\frac{(r-1) \left(l^2 r+l r-4\right)}{r^4} \right]\Bigg\}{\hat {\cal{Y}}}_{lm} (r) = 0. ~~~~\eqn
\end{widetext}
 Then, we expand $\hat {\cal{Y}}_{lm} (r)$ as
\bqn
\lb{expand2}
\hat {\cal{Y}}_{lm} (r) =  (r-1)^s \sum_{n=0}^\infty d_n \left(r- 1\right)^n,
\eqn
where $d_n$'s are constants to be determined with the Frobenius method. In particular,
inserting the above expression into Eq.(\ref{master2c}), we find two branches of solutions, given,  respectively,  by
\bq
\lb{expand2a}
s_{\pm} = \pm i \omega.
\eq
Inserting it together with Eq.(\ref{expand2}) into Eq.(\ref{master2c}) we find
\begin{widetext}
\bqn
\lb{eq4.9a}
 && n (n+2 i \omega ) d^+_n   +\left[ -l^2-l+2 n^2+n (-3+4 i \omega )+2 \omega ^2-3 i \omega +5 \right] d^+_{n-1} \nb\\
 && + \left[ -l^2-l+n^2+n (-3+2 i \omega )+5 \omega ^2-3 i \omega +2 \right] d^+_{n-2} + 4 \omega ^2 d^+_{n-3} + \omega ^2 d^+_{n-4}  = 0,
\eqn
for $s = s_+$, and
\bqn
\lb{eq4.9b}
 && n (n - 2 i \omega ) d^+_n   +\left[ -l^2-l+2 n^2+n (-3-4 i \omega )+2 \omega ^2+3 i \omega +5 \right] d^+_{n-1} \nb\\
 && + \left[ -l^2-l+n^2+n (-3-2 i \omega )+5 \omega ^2+3 i \omega +2 \right] d^+_{n-2} + 4 \omega ^2 d^+_{n-3} + \omega ^2 d^+_{n-4}  = 0,
\eqn
for $s = s_-$.
Then, similar to  $a_n$'s given by Eq.(\ref{eq4.7}), all these coefficients can be written   in terms of $d^{\pm}_0$. With the same  reasons as given for the choice of $a_0$ and $b_0$, without loss of the
generality, we can always set $d^{\pm}_0=1$. Therefore, for any given  $\omega$ and $l$, we find the general solution of $\hat {\cal{Y}}_{lm} (r)$ in the neighborhood of $r=1$,
\bqn
\lb{expand2b}
\hat {\cal{Y}}_{lm} (r) &=& (r-1)^{s_-} \sum_{n=0}^\infty d_n^- \left(r- 1\right)^n  +  {\cal{A}} (r-1)^{s_+} \sum_{n=0}^\infty d_n^+ \left(r- 1\right)^n,
\eqn
\end{widetext}
with $ {\cal{A}}$ being a complex number,   and $d_n^\pm$'s are given by Eqs.(\ref{eq4.9a}) and (\ref{eq4.9b}) with $d^{\pm}_0=1$.

It should be noted that in practice Eqs.~\eqref{expand1}, \eqref{expand3} and \eqref{expand2b} cannot be expanded to infinite orders. Instead, we shall expand them to the orders by imposing  certain  tolerances   {to the numerical errors} of our numerical computations, and make sure such orders are high enough, so that our main results and conclusions will remain the same,   {i.e., the numerical solutions adopt convergent behaviors}. 

 \subsection{Algorithm for Solving Master Equations}

 To  solve Eq.(\ref{master2}),  following  \cite{Chandra1975} we introduce the mode function $\Phi(r)$ by
\bqn
\lb{Phi2}
{\hat {\cal{Y}}}_{lm}(r) &=& \text{Exp}\left(i \int^{r} \Phi (r) d r\right),
\eqn
or equivalently
\bqn
\lb{Phi3}
\Phi(r) &=& - i \frac{d \ln {\hat {\cal{Y}}}_{lm}}{dr}.
\eqn
Then, Eq.(\ref{master2}) reduces to
\bqn
\lb{master3}
&& \frac{i (r-1)^2 }{r^2} \Phi'  -\frac{(r-1)^2 }{r^2} \Phi^2 +\frac{i \left(2 r^2-3 r+1\right) }{r^3} \Phi  \nb\\
&&  -\frac{(r-1) \left(l^2 r+l r-4\right)}{r^4}  +\omega ^2 =0. 
\eqn

It is remarkable to note that such introduced $\Phi(r)$ does not depend on the amplitudes of ${\hat {\cal{Y}}}_{lm}$ around any of the points $r = \left\{r_{\pm}, \; r_{\text{UH}}, \; r_{\text{max}}\right\}$, with $r_{\pm} \equiv 1 \pm \epsilon$ ( $0< \epsilon \ll 1$)  and $r_{\text{max}}$ a positive number to be specified. As a result, we can assign
  {$a_0, \; b_0$ and $d_0^\pm$ any values.} As mentioned previously, without loss of the generality, we shall set all of them to one.

For our convenience, we further introduce a new variable $z$ by $z \equiv 2 r/(r+3/4)$ in Eq.\eqref{master3}. In this way, $r \in [3/4, 1) \cup (1, \infty)$ is mapped to $z \in [1, 8/7) \cup (8/7, 2)$. Therefore, Eq.\eqref{master3} becomes
\bqn
\lb{master5}
&& \frac{2 i \left(7 z^2-22 z+16\right)^2}{27 z^2}  \frac{d \Phi}{d z}-\frac{(8-7 z)^2 }{9 z^2} \Phi^2 \nb\\
&& -\frac{8 i \left(35 z^3-138 z^2+168 z-64\right) }{27 z^3} \Phi  + \omega ^2 \nb\\
&& -\frac{16 (7 z-8) (z-2)^2 \left[ \left(3 l^2+3 l+16\right) z-32\right]}{81 z^4} = 0.\nb\\
\eqn
Thus, we obtain a complex ordinary differential equation (ODE) of the first order, which will be much easier to solve than  Eq.(\ref{master2}).
To solve Eq.(\ref{master5}) we use the shooting method. Specifically, for any given $l$, we integrate Eq.(\ref{master5}) as follows:

\begin{enumerate}
\item  With a good guess, we first choose a (complex) value of $\omega$. Then, by using the polynomial solution \eqref{expand3}, we create the ``initial value" ${\tilde \Psi}_{\text{UH}}(\omega) \equiv {\tilde \Psi}(r=r_{\text{UH}}, \omega)$. After that, we calculate $\Phi_{\text{UH}}(\omega) \equiv \Phi(z=z_{\text{UH}}, \omega)$  using Eq.\eqref{Phi3} {[and Eqs.\eqref{4.21a}, \eqref{master2ba}]}, with $z_{\text{UH}} = 2 r_{\text{UH}}/(r_{\text{UH}}+3/4)$.

\item With the same $\omega$, using the polynomial solution \eqref{expand1}, we create the ``initial value" $\Psi_{\text{max}}(\omega) \equiv \Psi(r=r_{\text{max}}, \omega)$. After that, we calculate $\Phi_{\text{max}}(\omega) \equiv \Phi(z=z_{\text{max}}, \omega)$  using Eq.\eqref{Phi3} {[and Eqs.\eqref{4.21a}, \eqref{Psi}]}, with $z_{\text{max}} = 2 r_{\text{max}}/(r_{\text{max}}+3/4)$. Here $r_{\text{max}}$ is chosen to be sufficiently large.

\item By using the initial value $\Phi_{\text{UH}}(\omega)$, integrate Eq.(\ref{master5}) from $z_{\text{UH}}$ to $z_{-} = r_{-}/(r_{-}+3/4)$ to obtain $\Phi_{\text{-}} \equiv \Phi(z_{\text{-}})$. Equalizing this $\Phi_{\text{-}}$ and its counterpart given by the polynomial \eqref{expand2b} {[and Eq.\eqref{Phi3}]}, we can solve for the combination coefficient ${\cal A}$, to be denoted by ${\cal A}_-$.

\item By using the initial value $\Phi_{\text{max}}(\omega)$, integrate Eq.(\ref{master5}) from $z_{\text{max}}$ to $z_{+} = r_{+}/(r_{+}+3/4)$ to obtain $\Phi_{\text{+}} \equiv \Phi(z_{\text{+}})$. Equalizing this $\Phi_{\text{+}}$ and its counterpart given by the polynomial \eqref{expand2b} {[and Eq.\eqref{Phi3}]}, we can again solve for a combination coefficient ${\cal A}$, to be denoted by ${\cal A}_+$.

\item  {Logically repeat steps 1 - 4} to search for the desired $\omega$, with which we have $\Delta {\cal A} \equiv |{\cal A}_- - {\cal A}_+| \lesssim \epsilon_{\text{max}}$, where $\epsilon_{\text{max}}$ represents our tolerance to the uncertainty. Clearly, this $\Delta {\cal A}$ could be seen as an indicator to the accuracy of our calculations.
Note that these iterations could be executed automatically by using {\it Mathematica}. %
\end{enumerate}

\begin{widetext}
\begin{table*}
\centering
	\caption{Usages of equations in generating the suitable boundary conditions for solving Eq.\eqref{master5}.}  
	\label{tableA}
	\begin{tabular}{|c  c | c c | c c| c| }
		\hline
		~Type of boundary condition ~  & \quad & ~ Characterized by ~ & \quad & ~  Polynomial ~  & \quad & ~Translated to $\Phi(r)$ with ~  \\
  	   & \quad & & \quad &  expression   & \quad &   \\
		\hline
    	\hline
		Pure out-going wave at $r \to +\infty$ &  &    ~Eq.\eqref{Asmptotic1}, with $a_-=0$ ~ & \quad &  Eq.\eqref{expand1}  & \quad &  Eqs.\eqref{Phi3}, \eqref{4.21a}, \eqref{Psi}
		\\
    	\hline
     		~Pure in-going wave at $r \to (3/4)^{+}$~ &  &    ~Eq.\eqref{Asmptotic2}, with $b_+=0$ ~ & \quad &  Eq.\eqref{expand3}  & \quad &  Eqs.\eqref{Phi3}, \eqref{4.21a}, \eqref{master2ba}
		\\
    	\hline
     		~Smoothness across $r = 1$ ~ &  &    N/A & \quad &  Eq.\eqref{expand2b}  & \quad &  Eq.\eqref{Phi3}
		\\
    	\hline
	\end{tabular}
\end{table*}
\end{widetext}

\subsection{Numerical Results}

Using the algorithm introduced in last subsection, we can obtain eigenvalues of  $\omega$ for any given $l$. 
 {However, since the algorithm introduced in this section is quite algebraically involved, before presenting our results, for the sake of readers' convenience, let us first  briefly provide a summary to the pivot equations we use in the shooting method. Basically, our main target is to obtain eigenvalues of $\omega$ by solving the first-order non-linear complex differential equation \eqref{master5}. For this purpose, suitable boundary conditions are needed. We summarize the relative necessary equations in Table \ref{tableA}. Eq.\eqref{master5} plus the ones given in Table \ref{tableA} form a complete system, and the eigenvalues of $\omega$ given in Table \ref{table1} are found by using the shooting method with the assistance of {\it Mathematica}.}  In this table we also provide the corresponding results from GR (see, e.g., \cite{Chao2022}). Comparing the results from  $\ae$-theory and GR, we can find the differences between these two theories.  
To characterize these discrepancies, let's define \cite{Berti2009, Ghosh2021}
\bqn
\lb{deltaw}
\delta f &\equiv& \frac{Re(\omega^{\text{\ae}})}{Re(\omega^{\text{GR}})} - 1, \nb\\
\delta \tau &\equiv& \frac{Im(\omega^{\text{GR}})}{Im(\omega^{\text{\ae}})} - 1,
\eqn
which are also given for different modes   in Tabel \ref{table1}. Using the current data provided by the LVK scientific collaboration \cite{Ghosh2021, LVK2022}, it is  hard to impose further constraints  on  $\ae$-theory. For instance, combining the results from  \cite{Ghosh2021, LVK2022}, we find that the most stringent constraint on $\delta f$ is right at the order $ {\cal {O}} (10^{-2})$, which is similar to the largest discrepancy that we have presented  in Table \ref{table1}.   Nonetheless,  such discrepancies should be well within the detectability of LISA, TianQin and Taiji. In particular, according to  Table II provided in \cite{Liu2020},    { the sensitivity of LISA and TianQin to, e.g., $\delta f$, could reach about $ {\cal {O}} (10^{-5})$.} Therefore, they are very likely to be able to distinguish the results from the $\ae$-theory and GR, as listed in Table \ref{table1}.

\begin{widetext}
\begin{table*}
\centering
	\caption{Values of $\omega$ for any given $l$  in $\ae$-theory and  GR, where we have set   {$c=G_N=r_{MH}(=r_s)=1$}.}  
	\label{table1}
	\begin{tabular}{|c c|c c|c c|c c|c c|c|}
		\hline
		$l$    & \quad &  $\ae$  & \quad &  $\Delta {\cal A} $  &  &  GR   &  &   $\delta f$   &  &  $\delta \tau$  \\
		\hline
    	\hline
		$l=2$
		&  &  ~$ -0.74314 	-0.17787 i$~  &  &   $4.76\times10^{-34}$  &  & ~$-0.74734 -0.17792  i$~ &  & -0.00562
   &  &  -0.00030
		\\
		&  &  ~$ -0.67267 	-0.54966 i$~  &  &  $4.54\times10^{-30}$   &  & $-0.69342 	-0.54783  i$ &  &  -0.02993
  &  & 0.00334
		\\
		&  &  ~$-0.56544 	-0.91304 i$~  &  &   $9.38\times10^{-27}$  &  & $-0.60211 	-0.95656  i$ &  &  -0.06090
   &  & -0.04549
		\\
    	\hline
    	\hline
		$l=3$
		&  &  ~$-1.19890 	-0.18506 i$~ &  &   $1.40\times10^{-33}$   &  & $-1.19889 	-0.18541 i$ &  &  0.00001
  &  & -0.00189

		\\
		&  &  ~$-1.16448 	-0.56037 i$~  &  &  $3.69\times10^{-30}$   &  & $-1.16529 	-0.56260 i$ &  & -0.00070
   &  & -0.00396

		\\
		&  &  ~$-1.07995 	-0.95446 i$~  &  &   $3.28\times10^{-27}$  &  & $-1.10337 	-0.95819 i$ &  &  -0.02122
  &  & -0.00388

		\\
    	\hline
    	\hline
		$l=4$
		&  &  ~$-1.61838 	-0.18832 i$~ &  &   $1.05\times10^{-18}$   &  & $-1.61836 	-0.18833 i$ &  &   0.00002
  &  & -0.00006

		\\
		&  &  ~$-1.59342 	-0.56847 i$~  &  &  $1.49\times10^{-29}$   &  & $-1.59326 	-0.56867 i$ &  &  0.00010
   &  & -0.00034

		\\
		&  &  ~$-1.54705 	-0.95738 i$~ &  &   $9.75\times10^{-25}$   &  & $-1.54542 	-0.95982 i$ &  &   0.00106
  &  & -0.00254

		\\
    	\hline
    	\hline
		$l=10$
		&  &  ~$-3.98329 	-0.57582 i$~ &  &   $2.14\times10^{-14}$   &  & $-3.98329 	-0.57582i$ &  &   0.00000
  &  & 0.00000
	\\
    	\hline
	\end{tabular}
\end{table*}

\end{widetext}


\section{ Instability of  the aether field }
\renewcommand{\theequation}{5.\arabic{equation}} \setcounter{equation}{0}

In the last section, we consider the metric perturbations only, represented by the master equation (\ref{master2}). In this section, we turn to study the aether perturbations given by
Eqs.(\ref{master4}) - (\ref{almvr}). To this goal, inserting Eq.\eqref{almvr} into Eq.\eqref{fieldeqn4}  we obtain
\bqn
\lb{masterofa}
\left\{ \kappa_1 \frac{d^2}{d r^2} + \kappa_2 \frac{d}{d r} + \kappa_3 \right\} {\hat {\cal{Z}}}_{lm} &=& 0,
\eqn
where $\kappa_{1, 2, 3}$   are given in Appendix A. Note that \eqref{masterofa} is at the same position as \eqref{master2c}. To solve it, following what we did in the last section, we first
specify its boundary conditions at the universal horizon and spatial infinity, as well as the smoothness conditions across the metric horizon, and then solve it with such conditions by the 
shooting method, one side from the UH to the MH, and the other side from $r_{\text{max}}$ to the MH.

\subsection{Boundary Conditions at the Universal Horizon and Spatial Infinity}

To find the proper boundary conditions for ${\hat {\cal{Z}}}_{lm} (r)$ at $r \to 3/4$ and $r \to +\infty$, we first note that in terms of $x$ Eq.\eqref{master4} takes the form,
\bqn
\lb{master6}
\left[   \frac{d^2}{d x^2} - V_{\text{eff}}(r) \right ] \psi &=& 0. ~~~~
\eqn
Note that \eqref{master6} is at the same position as \eqref{master2a}.

\subsubsection{At Spatial Infinity}

When $r \to +\infty$,  \eqref{master6} leads to $\psi = {\bar a}_+ + {\bar a}_- x$. For ${\hat {\cal{Z}}}_{lm}$ to be finite at $r\to +\infty$, we shall set ${\bar a}_-=0$ [cf. \eqref{psi}]. Thus,   we have
\bqn
\lb{expand1alm}
\psi &=&   \sum_{n=2}^\infty \frac{{\bar a}_n}{r^n},
\eqn
where ${\bar a}_n $ can be calculated through recursion relations similar to \eqref{eq4.7}. Note that it is known {\it a posteriori} that the polynomial solution of $\psi$  at $r \to +\infty$ starts from $n=2$,
as already  indicated in Eq.(\ref{expand1alm}). Combining with  \eqref{psi}, we have found the polynomial solution of ${\hat {\cal{Z}}}_{lm}$ in the neighborhood of $r \to +\infty$.

\subsubsection{At the Universal Horizon}

On the other hand, at the UH  {$r = 3/4$},   by applying the formulas presented in Appendix B, we can write \eqref{masterofa} in the form,
\bqn
\lb{master2balm}
 \left[{\bar W}_g(r) \frac{d^2}{d {\bar y}^2} + \left( -{\bar V}_g \right) \right] {\bar \psi} =0,
\eqn
where ${\bar W}_g  \equiv   \kappa_1/{\bar p}^2$, and
\begin{widetext}
\bqn
\lb{rast2alm}
{\bar \psi} &\equiv& - {\hat {\cal{Z}}}_{lm} \frac{\sqrt{\left(64 r^3-16 r^2-12 r-9\right) \left(2 \sqrt{2 l^2+2 l+9}+4 r-3\right)} }{r^2}\nb\\
&& \times \exp \left\{\frac{1}{8} i \omega  \left[3 \sqrt{6} \tanh ^{-1}\left(\frac{8 r+3}{3 \sqrt{8 r^2+4 r+\frac{3}{2}}}\right)-8 \tanh ^{-1}\left(\frac{20 r+7}{3 \sqrt{24 r (2 r+1)+9}}\right)\right]\right\},   \nb\\
\frac{dr}{d{\bar y}} & \equiv & {\bar p} (r) =  \frac{4 r-3}{2 \sqrt{2 l^2+2 l+9}+4 r-3},\nb\\
{\bar V}_g  &\equiv &  {- \left\{ \frac{{\bar W}_g}{4} \left[2 {\bar p} \frac{d \left({\bar p}'-{\bar p} \kappa_2/\kappa_1 \right)}{dr}  -\left({\bar p}'- \frac{ {\bar p} \kappa_2}{\kappa_1} \right)^2 \right] + \kappa_3. \right \}.}
\eqn
\end{widetext}
When $r \to (3/4)^+$, \eqref{master2balm} leads to $ {\bar \psi} = {\bar b}_+e^{+ {\bar y}} + {\bar b}_- e^{- {\bar y}}$. For ${\hat {\cal{Z}}}_{lm}$ to be finite at $r\to (3/4)^+$, we shall set ${\bar b}_-=0$ [cf. \eqref{psi}]. Thus, we have
\bqn
\lb{expand3alm}
{\bar \psi} &=&  e^{\bar y} \sum_{n=0}^\infty {\bar b}_n \left(r- \frac{3}{4} \right)^n.
\eqn
Combining with  \eqref{psi}, we have found the polynomial solution of ${\hat {\cal{Z}}}_{lm}$ in the neighborhood of $r \to (3/4)^+$.

\subsection{Smoothness Conditions Across the Metric Horizon}

At the MH ($r=1$), it is clear that  Eq.\eqref{masterofa} becomes singular. By following the same procedures of solving Eq.\eqref{master2c}, we can also solve
Eq.\eqref{masterofa}   in the neighborhood of $r=1$. To do so, we shall first write  ${\hat {\cal{Z}}}_{lm} (r)$ in the form,
\bqn
\lb{expand2alm}
{\hat {\cal{Z}}}_{lm} (r) =  (r-1)^{\bar s} \sum_{n=0}^\infty {\bar d}_n \left(r- 1\right)^n.
\eqn
Just like how we obtain \eqref{expand2b}, the solution is determined to be of the form
\begin{widetext}
\bqn
\lb{expand2balm}
{\hat {\cal{Z}}}_{lm} (r) &=& (r-1)^{-i \omega} \left[  \left(r- 1\right) \sum_{n=0}^\infty {\bar d}_n^- \left(r- 1\right)^n  +  {\cal{B}} \sum_{n=0}^\infty {\bar d}_n^+ \left(r- 1\right)^n \right],
\eqn
\end{widetext}
where ${\bar d}_n^{\pm}$ can be calculated through recursion relations similar to \eqref{eq4.9a} and \eqref{eq4.9b}, and the (complex) constant  ${\cal{B}}$ is acting as ${\cal{A}}$ in \eqref{expand2b}.

\subsection{Numerically Solving the aether  Master Equation}

To solve the aether master equation, let us first set
\bqn
\lb{Phi2alm}
{\hat {\cal{Z}}}_{lm}(r) &=& \text{Exp}\left(i \int^{r} \phi (r) d r\right),
\eqn
  {[It must not be confused  with $\phi$ introduced here and the one  appearing in Eq.\eqref{eq5.12aa}]} so that Eq.(\ref{masterofa}) takes the form,
\bq
\lb{master3alm}
\alpha_1 \phi'(r) + \alpha_2 \phi(r)^2 + \alpha_3\phi(r) + \alpha_4 = 0 ,
\eq
where $\alpha_n$'s are given in Eq.(\ref{A.7}) and must not be confused with    the PPN parameters mentioned in Sec. II.
With the above given boundary conditions at the universal horizon and spatial infinity, as well as the smoothness conditions across the metric horizon, following the steps presented in  Sec. IV.D,
 we can solve Eq.(\ref{master3alm})  with the shooting method  {As what we did in the last section, the new variable $z$ will be introduced in practice. In addition,  here in Table \ref{tableB} we provide a summary to the usages of those main equations that related to solving Eq.\eqref{master3alm}.}

 \begin{widetext}
\begin{table*}
\centering
	\caption{Usages of equations in generating the suitable boundary conditions for solving Eq.\eqref{master3alm}.}  
	\label{tableB}
	\begin{tabular}{|c  c | c  c| c| }
		\hline
		~Type of boundary condition ~ & \quad & ~  Polynomial ~  & \quad & ~Translated to $\phi(r)$ with ~  \\
  	   & \quad &  expression   & \quad &   \\
		\hline
    	\hline
		Finity  at $r \to +\infty$ &  \quad &  Eq.\eqref{expand1alm}  & \quad &  Eqs.\eqref{Phi2alm}, \eqref{psi}
		\\
    	\hline
     		~Finity at $r \to (3/4)^{+}$~ & \quad &  Eq.\eqref{expand3alm}  & \quad &  Eqs.\eqref{Phi2alm}, \eqref{rast2alm}
		\\
    	\hline
     		~Smoothness across $r = 1$ ~ &  \quad &  Eq.\eqref{expand2balm}  & \quad &  Eq.\eqref{Phi2alm}
		\\
    	\hline
	\end{tabular}
\end{table*}
\end{widetext}

However, different from what we have done in Sec. IV, in this section we would like to address the problem of instability found in \cite{TZZW21} for large $l$.
In particular,   in Table \ref{table2}, in which $\Delta{\cal{B}}$ is defined in the same manner as that of $\Delta{\cal{A}}$, we present the values of $\omega_{\ae}$ for $l=100$.  
In contrast to those given in Table \ref{table1},  now the imaginary parts of $\omega_{\ae}$'s are all positive,
which indicates the instability of the aether field. However,  we must note the differences between   {the tolerance levels of Table \ref{table1} indicated by  $\Delta {\cal A}$ and the ones given in Table \ref{table2},
 indicated by  $\Delta \cal B$.} In particular,   {the tolerance levels} given by $\Delta \cal B$ are much weaker than those given by $\Delta {\cal A}$.

\begin{table}
	\caption{Values of $\omega_{\ae}$ by solving Eq.(\ref{master3alm}) for the aether field  with $l=100$ and the units $c=G_N=r_{MH}=1$.}  
	\label{table2}
	\begin{tabular}{|c  c | c |}
		\hline
		$\omega_{\ae}$    & \quad & $\Delta {\cal B}$   \\
		\hline
    	\hline
		$-1944.19 +	87.4996i$	&  &  ~$ 1.16 \times 10^{-5} $~
		\\
	$-2913.80 +	155.310 i$	&  &  ~$ 5.19 \times 10^{-6} $~
		\\
$-9886.49 +	267.356 i$	&  &  ~$ 4.52 \times 10^{-7} $~
		\\
	$-139279.4  +	101836.6 i$ 	&  &  ~$ 1.13 \times 10^{-8} $~
		\\
	$-264002.5  +	102057.4 i$ 	&  &  ~$ 5.43 \times 10^{-9} $~
		\\
    	\hline
	\end{tabular}
\end{table}

In Table \ref{table3}, we also consider the cases $l = 2, 10$, from which it can be seen that unstable modes in each case are also identified, again with much weaker    {tolerance levels}
$\Delta {\cal B} \approx  {\cal O}\left(10^{-10}\right)$.


\begin{table}
	\caption{Values of $\omega_{\ae}$ by solving Eq.(\ref{master3alm}) for the aether field  with the units $c=G_N=r_{MH}=1$ for $l = 2, 10$, respectively.}  
	\label{table3}
	\begin{tabular}{|c|c|c|}
		\hline
		~$l$ ~  &   $\omega_{\ae}$  &    $\Delta {\cal B}$  \\
		\hline
    	\hline
		$2$ & ~ $ -43880 \pm 165.44 i$~ & ~  ${\cal O}\left(10^{-11}\right)~ $\\ \hline
		$10$ &    $-46071 \pm 291.48 i$ & ${\cal O}\left(10^{-10}\right)$\\
    	\hline
	\end{tabular}
\end{table}

\section{Conclusions}
\renewcommand{\theequation}{6.\arabic{equation}} \setcounter{equation}{0}

In this paper, we have investigated QNMs of black holes in Einstein-aether ($\ae$-) theory, in which  the gravitational sector
of the theory is described by three different species of gravitons with spins $0, 1, 2$, respectively \cite{JM01}. By properly choosing the four free dimensionless coupling constants, $c_i$'s, the theory is
 self-consistent, such as free of ghosts and kinetic  instability \cite{Jacobson}, and satisfies all the observations carried out so far \cite{OMW18,GHBBCYY,TZZW21}. However,
 to avoid the vacuum gravi-\v{C}erenkov radiation, such as cosmic rays,  each of these three species  must move with a speed that is  at least no less than the speed of light \cite{EMS05}. 

 Due to the existence of such superluminal motions, we expect that black holes in $\ae$-theory will be dramatically different from those in GR. In particular, the metric horizon that is used to define
 the boundary of a black hole in GR is no longer a one-way membrane, and particles with superluminal  speeds can escape to the spatial infinity, even they are initially trapped inside the metric 
 horizon. As a result, the inner boundary of a black hole now is defined by the universal horizon, a one-way membrane for particles with any  speeds, including the ones with infinitely large  ones
 \cite{BS11}. This in turn implies that at the metric horizon now both ``in-going" and ``out-going" radiations are expected to exist, which will alternate the QNM spectra of black holes. 
 
 In addition, due to the presence of the aether field, the uniqueness theorem of the Schwarzschild black hole does not hold any longer \cite{Jacobson}, and various black holes in $\ae$-theory exist  \cite{Chao2020b}, 
 which   can be still formed from gravitational collapse of realistic matter  \cite{Garfinkle2007}, including the formation of universal horizons \cite{BMSW18}.

The change of the location of a black hole inner boundary from the metric horizon to the universal horizon also results in several technical challenges in the studies of QNMs of black holes. First,
 the notion of ``in-going" and ``out-going" radiations must be define properly, as now the universal horizon is always inside the metric horizon, in which the Schwarzschild timelike coordinate becomes
 spacelike. In addition, across the metric horizon, the perturbation equations are singular, and proper smoothness conditions must be imposed, too, before solving the corresponding boundary conditions
 imposed at the universal horizon and the spatial infinity.  In GR, the  field equations of perturbations are also singular at 
 a metric horizon,  but the problem is  nicely avoided, as now it is the inner boundary, and the smoothness conditions are simply replaced by the inner boundary conditions.

To overcome the above problems,   in this paper we have studied  the QNMs  of the Schwarzschild black hole, which is also a solution of  the Einstein-aether theory \cite{Chao2020b}, in which the presence of
 the   {aether field}  has no effects on  the spacetime, although the aether field is non-trivial [cf. Eqs.(\ref{anasolu1}) and (\ref{eq3.0})]. Despite the fact that the background is the same as that in GR, it is expected 
 that the corresponding QNMs are different, as   now the inner boundary is moved to the universal horizon, and both ``in-going" and ``out-going" radiations exist at the metric horizon. In addition, 
 the perturbation equations are also different [cf. Eqs.(\ref{fieldeqn4}) and (\ref{master1})].

In particular, in  Section III we have constructed the two master equations (\ref{master2a}) and (\ref{master4}), respectively, for the metric and aether field perturbations, while in Section IV we have first worked out step by step the boundary conditions for the master equation of the metric perturbations at the universal horizon, $r_{UH} = 3r_s/4$, where $r_s = 2m$ is the location of the metric horizon, and at the spatial infinity. Then, we have
systematically worked out  its  
smoothness conditions across the metric horizon. Afterwards, combining the Chandrasekhar-Detweiler method \cite{Chandra1975} with the shooting one, one side  is from the universal horizon to the metric horizon, while the other side  is  from the spatial infinity to the metric horizon [cf. Section IV.D], we have found various modes $\omega$ for $l = 2, 3, 4, 10$, given explicitly in Table \ref{table1}. In the same table we have also presented the corresponding
values of the modes in GR, and found that the differences between the ones given in  $\ae$-theory and those given in GR {can hardly be distinguished} by current observations of gravitational waves, although it is well within 
the detectability of the detectors of the next generation. 

In Section V, following closely what we have done  for the metric perturbations in Section IV,  we have considered the aether perturbations and found various modes  $\omega_{\ae}$   {[cf. Eq.\eqref{almvr}]} for $l = 2, 10, 100$, given,
respectively in Tables \ref{table2} and \ref{table3}. In these two tables, in contrast to Table \ref{table1},  we have identified several modes of $\omega_{\ae}$'s, which all have  a positive imaginary part. This strongly
indicates the instability of the aether field. This is also consistent with what was found in \cite{TZZW21} for the  Schwarzschild black hole against large angular perturbations. However, due to our numerical errors of the current paper, it must be noted that such a conclusion is not exclusive. In particular, in the metric perturbations presented in  Section IV, we have not been able to find such an instability.

Therefore, it would be very important to confirm the above instability,  { for example, by considering the even-parity perturbations and meanwhile  extending the current studies to other black holes found in \cite{Chao2020b}.  
It would be also very interesting to confirm the isospectrality  of the theory \cite{Chao2022}.
However, due to the scope of this paper, we would like to address these important questions in other occasions. }

\section*{Acknowledgments}

We would like to express our gratitude  to Prof. Shinji Tsujikawa for his valuable comments and suggestions, and to Dr. Xiang Zhao for  the early stage of his collaboration of this project. This work is supported in part by the National Key Research and Development Program of China Grant No. 2020YFC2201503, the National Natural Science Foundation of China under Grant No. 12205254, 12275238, 11975203 and 11675143,  the Zhejiang Provincial Natural Science Foundation of China under Grant No. LR21A050001 and LY20A050002, and the Fundamental Research Funds for the Provincial Universities of Zhejiang in China under Grant No. RF-A2019015.


\section*{Appendix A: The coefficients of $\rho_{abc}$, $\kappa_{n}$ and $\alpha_m$}
\renewcommand{\theequation}{A. \arabic{equation}} \setcounter{equation}{0}

With the gauge condition $C_{lm} = 0$, the linearized Einstein-aether field equations 
\bq
\lb{A.1}
E_{\phi t}=E_{\phi r}=E_{\phi \theta}=\AE^{\phi}=0,
\eq
 can be cast respectively in the forms,
\begin{widetext}
\bqn
\lb{fieldeqn1}
&& \rho_{101} {\ddot J}_{lm} + \rho_{102} {\dot J}_{lm} + \rho_{103} {\dot J}_{lm}'  + \rho_{104} a_{lm} +\rho_{105} {\ddot a}_{lm} +  \rho_{106} {\dot a}_{lm}  + \rho_{107} {\dot a}_{lm}' + \rho_{108} a_{lm}'+ \rho_{109} a_{lm}''+ \rho_{110} {\dot G}_{lm}= 0, ~~~~~~\\
\lb{fieldeqn2}
&& \rho_{201} {J}_{lm} + \rho_{202} {\dot J}_{lm} + \rho_{203} {\dot J}_{lm}'   + \rho_{204} a_{lm} +\rho_{205} {\ddot a}_{lm} +  \rho_{206} {\dot a}_{lm}   + \rho_{207} {\dot a}_{lm}' + \rho_{208} a_{lm}'+ \rho_{209} a_{lm}''  + \rho_{210} {G}_{lm} \nb\\
&& ~~~~~~~~~~~~~~~~~~~~~~~~~~~~~~~~~~~~~~~~~~~~~~~~~~~~~~~~~~~~~~~~~~~~~~~~~~~~~~~~~~~~~~~~~~~~~~~~~~~~~~~~~~~~~~~~~~~~~~~+ \rho_{211} { G}_{lm}' = 0,~~~~~~~~\\
\lb{fieldeqn3}
&& \rho_{301} {J}_{lm} + \rho_{302} {\dot J}_{lm} + \rho_{303} {J}_{lm}'   + \rho_{304} G_{lm} +\rho_{305} {\dot G}_{lm} + \rho_{306} {\dot G}_{lm}'   + \rho_{307} G_{lm}'+ \rho_{308} G_{lm}'' = 0,~~~~~~~~\\
\lb{fieldeqn4}
&& \rho_{401} a_{lm} +\rho_{402} {\ddot a}_{lm} +\rho_{403} {\dot a}_{lm}  + \rho_{404} {\dot a}_{lm}' + \rho_{405} a_{lm}'+ \rho_{406} a_{lm}'' = 0,~~~~~~~~
\eqn
where 
 \bqn  
\lb{rho101}
\rho_{101} &=& -\frac{1}{2},\nb\\
\rho_{102} &=& \frac{1-r}{r^2} ,\nb\\
\rho_{103} &=& \frac{1}{2} \left(\frac{1}{r}-1\right),\nb\\
\rho_{104} &=&  \left\{256 r^8 \left(16 r^2+8 r+3\right) \sqrt{256 (r-1) r^3+27} \left[\sqrt{256 (r-1) r^3+27}-3 \sqrt{3}\right]^5 \right \}^{-1} \nb\\
&& \times \Big \{ c_1 \Big[98304 \sqrt{3} r^8-196608 \sqrt{3} r^7+98304 \sqrt{3} r^6+31104 \sqrt{3} r^4-31104 \sqrt{3} r^3+6912 \sqrt{256 (r-1) r^3+27} r^3 \nb\\
&& -729 \sqrt{256 (r-1) r^3+27}-8192 \sqrt{256 (r-1) r^3+27} r^8+16384 \sqrt{256 (r-1) r^3+27} r^7 \nb\\
&& -8192 \sqrt{256 (r-1) r^3+27} r^6-6912 \sqrt{256 (r-1) r^3+27} r^4+2187 \sqrt{3}\Big] \nb\\
&& \times \left[-256 r^4+256 r^3+3 \sqrt{768 (r-1) r^3+81}-27\right] \nb\\
&& \times \left[ 1024 \left(l^2+l-2\right) r^5+128 \left(3 l^2+3 l-4\right) r^4+2048 l (l+1) r^6+384 r^3+3024 r^2+1512 r+567\right] \Big \},\nb\\
\rho_{105} &=& -\frac{c_1 \left[ 3 \sqrt{3}-\sqrt{256 (r-1) r^3+27}\right] \left[ -256 r^4+256 r^3+3 \sqrt{768 (r-1) r^3+81}-27\right]}{4096 (r-1)^2 r^4},\nb\\
\rho_{106} &=& \frac{3 c_1 (4 r-3) \left[\sqrt{768 (r-1) r^3+81}-9\right] \left[-256 r^4+256 r^3+3 \sqrt{768 (r-1) r^3+81}-27\right]}{8192 (r-1)^2 r^6 \sqrt{256 (r-1) r^3+27}} ,\nb\\
\rho_{107} &=&  \frac{c_1 \left[ 256 r^4-256 r^3-3 \sqrt{768 (r-1) r^3+81}+27\right]^2}{2048 (r-1) r^5 \left[ 3 \sqrt{3}-\sqrt{256 (r-1) r^3+27}\right]} ,\nb\\
\rho_{108} &=& \Big\{ 512 r^7 \sqrt{256 (r-1) r^3+27} \left[\sqrt{256 (r-1) r^3+27}-3 \sqrt{3}\right]^4 \Big\}^{-1} \nb\\
&& \times c_1 \left(64 r^3-27\right) \left[ 256 r^4-256 r^3-3 \sqrt{768 (r-1) r^3+81}+27\right]^2 \nb\\
&& \times \left[ -128 r^4+128 r^3+3 \sqrt{768 (r-1) r^3+81}-27\right] ,\nb\\
\rho_{109} &=& -\frac{c_1 \left[ -256 r^4+256 r^3+3 \sqrt{768 (r-1) r^3+81}-27\right]^3}{4096 r^6 \left[ 3 \sqrt{3}-\sqrt{256 (r-1) r^3+27}\right]^3},\nb\\
\rho_{110} &=& -\frac{l^2+l-2}{4 r^2},
\eqn
 \bqn
\lb{rho201}
\rho_{201} &=& -\frac{l^2+l-2}{2 r^2}, \nb\\
\rho_{202} &=& \frac{1}{r}, \nb\\
\rho_{203} &=& \frac{1}{2}, \nb\\
\rho_{204} &=& \Big\{ 256 (r-1) r^7 \left(16 r^2+8 r+3\right) \sqrt{256 (r-1) r^3+27} \left[ \sqrt{256 (r-1) r^3+27}-3 \sqrt{3}\right]^3 \Big\}^{-1} \nb\\
&& \times c_1 \Big[ 98304 \sqrt{3} r^8-196608 \sqrt{3} r^7+98304 \sqrt{3} r^6+31104 \sqrt{3} r^4-31104 \sqrt{3} r^3+6912 \sqrt{256 (r-1) r^3+27} r^3 \nb\\
&& -729 \sqrt{256 (r-1) r^3+27}-8192 \sqrt{256 (r-1) r^3+27} r^8+16384 \sqrt{256 (r-1) r^3+27} r^7 \nb\\
&& -8192 \sqrt{256 (r-1) r^3+27} r^6-6912 \sqrt{256 (r-1) r^3+27} r^4+2187 \sqrt{3}\Big] \nb\\
&& \times \left[1024 \left(l^2+l-2\right) r^5+128 \left(3 l^2+3 l-4\right) r^4+2048 l (l+1) r^6+384 r^3+3024 r^2+1512 r+567\right], \nb\\
\rho_{205} &=& -\frac{c_1 \left[ 3 \sqrt{3}-\sqrt{256 (r-1) r^3+27}\right]^3}{4096 (r-1)^3 r^3}, \nb\\
\rho_{206} &=& \frac{3 c_1 (4 r-3) \left[\sqrt{256 (r-1) r^3+27}-3 \sqrt{3}\right]^2 \left[ \sqrt{768 (r-1) r^3+81}-9\right]}{8192 (r-1)^3 r^5 \sqrt{256 (r-1) r^3+27}} , \nb\\
\rho_{207} &=&  \frac{c_1 \left[3 \sqrt{3}-\sqrt{256 (r-1) r^3+27}\right] \left[ -256 r^4+256 r^3+3 \sqrt{768 (r-1) r^3+81}-27\right]}{2048 (r-1)^2 r^4}, \nb\\
\rho_{208} &=& \frac{c_1 \left(64 r^3-27\right) \left(256 r^4-256 r^3-3 \sqrt{768 r^4-768 r^3+81}+27\right)}{1024 (r-1) r^6 \sqrt{256 r^4-256 r^3+27}}, \nb\\
\rho_{209} &=& -\frac{c_1 \left[ 256 r^4-256 r^3-3 \sqrt{768 (r-1) r^3+81}+27\right]^2}{4096 (r-1) r^5 \left[ 3 \sqrt{3}-\sqrt{256 (r-1) r^3+27}\right]}, \nb\\
\rho_{210} &=& \frac{l^2+l-2}{2 r^3}, \nb\\
\rho_{211} &=& -\frac{l^2+l-2}{4 r^2}, \\
\lb{rho301}
\rho_{301} &=& \frac{1}{2 r^2}, \nb\\
\rho_{302} &=& \frac{1}{2}, \nb\\
\rho_{303} &=& \frac{r-1}{2 r}, \nb\\
\rho_{304} &=& \frac{r-2}{2 r^3}, \nb\\
\rho_{305} &=& -\frac{1}{2 r},\nb\\
\rho_{306} &=& \frac{1}{2}, \nb\\
\rho_{307} &=& \frac{3-2 r}{4 r^2}, \nb\\
\rho_{308} =&=&\frac{r-1}{4 r},
\eqn
 \bqn
\lb{rho401}
\rho_{401} &=&  \Big\{ 16 r^8 \left(16 r^2+8 r+3\right) \sqrt{256 (r-1) r^3+27} \left[ \sqrt{256 (r-1) r^3+27}-3 \sqrt{3}\right]^4 \Big\}^{-1} \nb\\
&& \times \Big\{ \Big[-98304 \sqrt{3} r^8+196608 \sqrt{3} r^7-98304 \sqrt{3} r^6-31104 \sqrt{3} r^4+31104 \sqrt{3} r^3-6912 \sqrt{256 (r-1) r^3+27} r^3 \nb\\
&& +729 \sqrt{256 (r-1) r^3+27}+8192 \sqrt{256 (r-1) r^3+27} r^8-16384 \sqrt{256 (r-1) r^3+27} r^7 \nb\\
&& +8192 \sqrt{256 (r-1) r^3+27} r^6+6912 \sqrt{256 (r-1) r^3+27} r^4-2187 \sqrt{3}\Big] \nb\\
&& \times \left[1024 \left(l^2+l-2\right) r^5+128 \left(3 l^2+3 l-4\right) r^4+2048 l (l+1) r^6+384 r^3+3024 r^2+1512 r+567\right] \Big\}, \nb\\
\rho_{402} &=&  -\frac{\left(\sqrt{256 (r-1) r^3+27}-3 \sqrt{3}\right)^2}{256 (r-1)^2 r^4}, \nb\\
\rho_{403} &=&  -\frac{3 (4 r-3) \left[\sqrt{256 (r-1) r^3+27}-3 \sqrt{3}\right] \left[\sqrt{768 (r-1) r^3+81}-9\right]}{512 (r-1)^2 r^6 \sqrt{256 (r-1) r^3+27}}, \nb\\
\rho_{404} &=& \frac{-256 r^4+256 r^3+3 \left(\sqrt{768 r^4-768 r^3+81}-9\right)}{128 (r-1) r^5} , \nb\\
\rho_{405} &=&  -\frac{\left(64 r^3-27\right) \left[128 r^4-128 r^3-3 \sqrt{768 (r-1) r^3+81}+27\right] \left[ 256 r^4-256 r^3-3 \sqrt{768 (r-1) r^3+81}+27\right]}{32 r^7 \sqrt{256 (r-1) r^3+27} \left(\sqrt{256 (r-1) r^3+27}-3 \sqrt{3}\right)^3}, \nb\\
\rho_{406} &=& -\frac{27}{256 r^6}+\frac{1}{r^3}-\frac{1}{r^2}.
\eqn

It is interesting to note that the coefficients $\rho_{abc}$ are functions of  $r$, $c_1$ and $l$ only, and some of them contain no $c_1$. 


The coefficients $\kappa_{n}$ appearing in Eq.(\ref{masterofa}) are given by
\bqn
\lb{kappa12}
\kappa_0 &\equiv& \sqrt{256 (r-1) r^3+27}, \nb\\
 \kappa_1 &\equiv& \frac{(r-1)^2}{r^2}, \nb\\
\kappa_2 &\equiv&  \Big[ 4194304 r^{14}-9437184 r^{13}+5505024 r^{12}-65536 r^{11}+2916 \left(\sqrt{768 (r-1) r^3+81}-9\right) r^3 \nb\\
&& -49152 \left(3 \sqrt{768 (r-1) r^3+81}-53\right) r^{10}+36864 \left(5 \sqrt{768 (r-1) r^3+81}-91\right) r^9 \nb\\
&& -9216 \left(\sqrt{768 (r-1) r^3+81}-15\right) r^8-6912 \left(\sqrt{768 (r-1) r^3+81}-15\right) r^7 \nb\\
&& -41472 \left(\sqrt{768 (r-1) r^3+81}-12\right) r^6+5184 \left(\sqrt{768 (r-1) r^3+81}-9\right) r^5 \nb\\
&& +3888 \left(\sqrt{768 (r-1) r^3+81}-9\right) r^4  \Big]^{-1} \nb\\
&& \times \Big\{ \Big[ 4194304 r^{12}-13631488 r^{11}+14942208 r^{10}-7340032 r^9+15552 \left(7 \sqrt{768 (r-1) r^3+81}-95\right) r^3 \nb\\
&& +7776 \left(\sqrt{768 (r-1) r^3+81}-9\right) r-11664 \left(\sqrt{768 (r-1) r^3+81}-9\right) \nb\\
&& -16384 \left(9 \sqrt{768 (r-1) r^3+81}-487\right) r^8+12288 \left(27 \sqrt{768 (r-1) r^3+81}-917\right) r^7 \nb\\
&& -9216 \left(21 \sqrt{768 (r-1) r^3+81}-571\right) r^6+64512 \left(\sqrt{768 (r-1) r^3+81}-15\right) r^5 \nb\\
&& -6912 \left(23 \sqrt{768 (r-1) r^3+81}-327\right) r^4-1296 \left(\sqrt{768 (r-1) r^3+81}-9\right) r^2 \Big] \nb\\
&& +\Big[ 1944 i \left(\sqrt{768 (r-1) r^3+81}-9\right) r^3+98304 i \left(\sqrt{768 (r-1) r^3+81}-27\right) r^{10} \nb\\
&& -221184 i \left(\sqrt{768 (r-1) r^3+81}-27\right) r^9+129024 i \left(\sqrt{768 (r-1) r^3+81}-27\right) r^8 \nb\\
&& -1536 i \left(\sqrt{768 (r-1) r^3+81}-27\right) r^7+4608 i \left(11 \sqrt{768 (r-1) r^3+81}-135\right) r^6 \nb\\
&& -3456 i \left(19 \sqrt{768 (r-1) r^3+81}-243\right) r^5+2592 i \left(\sqrt{768 (r-1) r^3+81}-9\right) r^4 \nb\\
&& +5832 i \left(\sqrt{768 (r-1) r^3+81}-9\right) r^2\Big] \omega_{\ae}\Big\},
\eqn
\bqn
\lb{kappa3}
\kappa_3 &\equiv&  \Big[ -268435456 \left(12 \sqrt{3}-\sqrt{256 (r-1) r^3+27}\right) r^{18}+671088640 \left(12 \sqrt{3}-\sqrt{256 (r-1) r^3+27}\right) r^{17} \nb\\
&& -452984832 \left(12 \sqrt{3}-\sqrt{256 (r-1) r^3+27}\right) r^{16}+16777216 \left(12 \sqrt{3}-\sqrt{256 (r-1) r^3+27}\right) r^{15} \nb\\
&& -1048576 \left(1488 \sqrt{3}-259 \sqrt{256 (r-1) r^3+27}\right) r^{14}+7864320 \left(336 \sqrt{3}-55 \sqrt{256 (r-1) r^3+27}\right) r^{13} \nb\\
&& -15925248 \left(16 \sqrt{3}-3 \sqrt{256 (r-1) r^3+27}\right) r^{12}-10616832 \left(16 \sqrt{3}-3 \sqrt{256 (r-1) r^3+27}\right) r^{11} \nb\\
&& -3981312 \left(109 \sqrt{3}-24 \sqrt{256 (r-1) r^3+27}\right) r^{10}+5971968 \left(15 \sqrt{3}-4 \sqrt{256 (r-1) r^3+27}\right) r^9 \nb\\
&& +3732480 \left(15 \sqrt{3}-4 \sqrt{256 (r-1) r^3+27}\right) r^8+2239488 \left(15 \sqrt{3}-4 \sqrt{256 (r-1) r^3+27}\right) r^7 \nb\\
&& -2519424 \left(3 \sqrt{3}-\sqrt{256 (r-1) r^3+27}\right) r^6-1259712 \left(3 \sqrt{3}-\sqrt{256 (r-1) r^3+27}\right) r^5 \nb\\
&& -472392 \left(3 \sqrt{3}-\sqrt{256 (r-1) r^3+27}\right) r^4  \Big]^{-1} \nb\\
&& \times \Big\{ \Big[  -268435456 \left(12 \sqrt{3}-\kappa_0\right) r^{15}+1006632960 \left(12 \sqrt{3}-\kappa_0\right) r^{14}-1291845632 \left(12 \sqrt{3}-\kappa_0\right) r^{13} \nb\\
&& +866123776 \left(12 \sqrt{3}-\kappa_0\right) r^{12}-9437184 \left(1488 \sqrt{3}-139 \kappa_0\right) r^{11}+1441792 \left(13524 \sqrt{3}-1397 \kappa_0\right) r^{10} \nb\\
&& -786432 \left(12489 \sqrt{3}-1412 \kappa_0\right) r^9+2654208 \left(671 \sqrt{3}-118 \kappa_0\right) r^8-1327104 \left(3161 \sqrt{3}-584 \kappa_0\right) r^7 \nb\\
&& +497664 \left(6763 \sqrt{3}-1282 \kappa_0\right) r^6-1492992 \left(78 \sqrt{3}-19 \kappa_0\right) r^5+186624 \left(831 \sqrt{3}-221 \kappa_0\right) r^4 \nb\\
&& -279936 \left(1527 \sqrt{3}-397 \kappa_0\right) r^3+6613488 \left(3 \sqrt{3}-\kappa_0\right) r^2+4408992 \left(3 \sqrt{3}-\kappa_0\right) r \Big] \nb\\
&& +\left[28311552 \left(12 \sqrt{3}-\kappa_0\right) r^{14}-42467328 \left(12 \sqrt{3}-\kappa_0\right) r^{13}+5308416 \left(12 \sqrt{3}-\kappa_0\right) r^{12} \right. \nb\\
&& \left. +3538944 \left(12 \sqrt{3}-\kappa_0\right) r^{11}+1327104 \left(129 \sqrt{3}-22 \kappa_0\right) r^{10}-5971968 \left(9 \sqrt{3}-2 \kappa_0\right) r^9  \right. \nb\\
&& \left. -3732480 \left(9 \sqrt{3}-2 \kappa_0\right) r^8-2239488 \left(9 \sqrt{3}-2 \kappa_0\right) r^7+2519424 \left(3 \sqrt{3}-\kappa_0\right) r^6+1259712 \left(3 \sqrt{3}-\kappa_0\right) r^5  \right. \nb\\
&& \left. +472392 \left(3 \sqrt{3}-\kappa_0\right) r^4\right] \omega_{\ae}^2  +6613488 \left(3 \sqrt{3}-\kappa_0\right) \nb\\
&& +l^2 \Big[ 268435456 \left(12 \sqrt{3}-\kappa_0\right) r^{16}-939524096 \left(12 \sqrt{3}-\kappa_0\right) r^{15}+1124073472 \left(12 \sqrt{3}-\kappa_0\right) r^{14} \nb\\
&& -469762048 \left(12 \sqrt{3}-\kappa_0\right) r^{13}+4194304 \left(339 \sqrt{3}-62 \kappa_0\right) r^{12}-2097152 \left(1599 \sqrt{3}-302 \kappa_0\right) r^{11} \nb\\
&& -+3932160 \left(591 \sqrt{3}-110 \kappa_0\right) r^{10}-7077888 \left(9 \sqrt{3}-2 \kappa_0\right) r^9+884736 \left(153 \sqrt{3}-43 \kappa_0\right) r^8 \nb\\
&& -1327104 \left(225 \sqrt{3}-59 \kappa_0\right) r^7+4478976 \left(3 \sqrt{3}-\kappa_0\right) r^6+2985984 \left(3 \sqrt{3}-\kappa_0\right) r^5+4478976 \left(3 \sqrt{3}-\kappa_0\right) r^4 \Big] \nb\\
&& +l \Big[ 268435456 \left(12 \sqrt{3}-\kappa_0\right) r^{16}-939524096 \left(12 \sqrt{3}-\kappa_0\right) r^{15}+1124073472 \left(12 \sqrt{3}-\kappa_0\right) r^{14} \nb\\
&& -469762048 \left(12 \sqrt{3}-\kappa_0\right) r^{13}+4194304 \left(339 \sqrt{3}-62 \kappa_0\right) r^{12}-2097152 \left(1599 \sqrt{3}-302 \kappa_0\right) r^{11} \nb\\
&& +3932160 \left(591 \sqrt{3}-110 \kappa_0\right) r^{10}-7077888 \left(9 \sqrt{3}-2 \kappa_0\right) r^9+884736 \left(153 \sqrt{3}-43 \kappa_0\right) r^8 \nb\\
&& -1327104 \left(225 \sqrt{3}-59 \kappa_0\right) r^7+4478976 \left(3 \sqrt{3}-\kappa_0\right) r^6+2985984 \left(3 \sqrt{3}-\kappa_0\right) r^5+4478976 \left(3 \sqrt{3}-\kappa_0\right) r^4\Big] \nb\\
&& +\Big[ -1610612736 i \sqrt{3} r^{17}+5234491392 i \sqrt{3} r^{16}-5737807872 i \sqrt{3} r^{15}+2139095040 i \sqrt{3} r^{14} \nb\\
&& -12582912 i \left(149 \sqrt{3}-18 \kappa_0\right) r^{13}+9437184 i \left(445 \sqrt{3}-54 \kappa_0\right) r^{12}-7077888 i \left(347 \sqrt{3}-42 \kappa_0\right) r^{11} \nb\\
&& +1769472 i \left(15 \sqrt{3}-2 \kappa_0\right) r^{10}-5308416 i \left(111 \sqrt{3}-22 \kappa_0\right) r^9+3981312 i \left(195 \sqrt{3}-38 \kappa_0\right) r^8 \nb\\
&& -2985984 i \left(9 \sqrt{3}-2 \kappa_0\right) r^7-2239488 i \left(9 \sqrt{3}-2 \kappa_0\right) r^6-3359232 i \left(27 \sqrt{3}-7 \kappa_0\right) r^5 \nb\\
&& +2519424 i \left(3 \sqrt{3}-\kappa_0\right) r^4+1889568 i \left(3 \sqrt{3}-\kappa_0\right) r^3+1417176 i \left(3 \sqrt{3}-\kappa_0\right) r^2 \Big] \omega_{\ae}\Big\}.
\eqn

The coefficients $\alpha_{n}$ appearing in Eq.(\ref{master3alm}) are given by
\bqn
\lb{A.7}
\alpha_1 &=& \frac{i (r-1)^2}{r^2}, \nb\\
\alpha_2 &=&  -\frac{(r-1)^2}{r^2}, \nb\\
\alpha_3&=&  -\frac{6 (r-1) \sqrt{768 r^4-768 r^3+81} \omega _{\text{\ae}}}{256 r^5-256 r^4+27 r} \nb\\
&& +\frac{16 i (r-1)^2 \left(16 r^2+12 r+9\right)}{r^3 (4 r-3) \left(16 r^2+8 r+3\right) \sqrt{256 (r-1) r^3+27} \left[\sqrt{256 (r-1) r^3+27}-3 \sqrt{3}\right]^3} \nb\\
&& \times  \Big[ 16384 r^8-32768 r^7+16384 r^6+8640 r^4+576 \sqrt{768 (r-1) r^3+81} r^3-8640 r^3 \nb\\
&& ~~~ -81 \sqrt{768 (r-1) r^3+81}-576 \sqrt{768 (r-1) r^3+81} r^4+729 \Big], \nb\\
\alpha_4 &=& -\frac{27 \omega _{\text{\ae}}^2}{256 r^4-256 r^3+27} \nb\\
&& -\frac{12 i \left(128 r^4-128 r^3+27\right) \omega _{\text{\ae}}}{r^2 (4 r-3) \left(16 r^2+8 r+3\right) \sqrt{256 (r-1) r^3+27} \left(\sqrt{256 (r-1) r^3+27}-3 \sqrt{3}\right)^3} \nb\\
&& \times  \left[-1728 r^4-64 \sqrt{3} \kappa_0 r^3+1728 r^3+27 \sqrt{3} \kappa_0 +64 \sqrt{3} \kappa_0 r^4-243\right] \nb\\
&& +\frac{16 (r-1)^2 \left(1024 \left(l^2+l-2\right) r^5+128 \left(3 l^2+3 l-4\right) r^4+2048 l (l+1) r^6+384 r^3+3024 r^2+1512 r+567\right)}{(3-4 r)^2 r^4 \left(16 r^2+8 r+3\right)^2 \sqrt{256 (r-1) r^3+27} \left(\sqrt{256 (r-1) r^3+27}-3 \sqrt{3}\right)^4} \nb\\
&& \times \Big(98304 \sqrt{3} r^8-196608 \sqrt{3} r^7+98304 \sqrt{3} r^6+31104 \sqrt{3} r^4-31104 \sqrt{3} r^3+6912 \kappa_0 r^3-729 \kappa_0-8192 \kappa_0 r^8 \nb\\
&& ~~~ +16384 \kappa_0 r^7-8192 \kappa_0 r^6-6912 \kappa_0 r^4+2187 \sqrt{3} \Big) .
\eqn
\end{widetext}


\section*{Appendix B: The formulas used in modifying a master equation}
\renewcommand{\theequation}{B. \arabic{equation}} \setcounter{equation}{0}

Suppose we have an eqaution of the form
\bqn
\lb{Beqn1}
\left[\eta_1(r) \frac{d^2}{d r^2} +\eta_2(r) \frac{d}{dr} +\eta_3(r)  \right] J(r) = 0.
\eqn
Let us introduce a new function $\Psi$ and a new variable $x$ by
\bqn
\lb{Beqn2}
J = q(r) \Psi, \quad \frac{dr}{dx} = p(r).
\eqn
Then, choosing  $q(r)$ as
\bqn
\lb{Beqn3}
q &=& \sqrt{p(r)} \exp{\left( -\frac{1}{2} \int \frac{\eta_2}{\eta_1} dr \right)}, 
\eqn
we find that  Eq.\eqref{Beqn1} can be written in a form of the Schr{\"o}dinger-like differential equation
\bqn
\lb{Beqn4}
\left[\frac{\eta_1}{p^2} \frac{d^2}{d x^2} - V(r)  \right] \Psi = 0,
\eqn
where
\bqn
\lb{Beqn5}
V(r) &\equiv& -\Bigg\{ \frac{\eta_1}{4 p^2} \bigg[ 2 p \frac{d\left( p'-p \eta_2/\eta_1 \right)}{dr}  \nb\\
&& ~~~~~~ - \left( p'-p \frac{\eta_2}{\eta_1} \right)^2 \bigg] + \eta_3 \Bigg\}, 
\eqn
and a prime denotes the derivative with respect to $r$. 
Note that another useful form of Eq.\eqref{Beqn4} is given by
\bqn
\lb{Beqn6}
\left[\frac{\eta_1}{p^2} \left(p^2 \frac{d^2}{d r^2} + p p' \frac{d}{d r} \right) - V(r)  \right] \Psi = 0,
\eqn
where Eq.\eqref{Beqn2} has been used. In addition, a common choice of $p(r)$ is   $p(r)=\sqrt{\eta_1}$.


\end{document}